\documentclass[conference]{IEEEtran}
\IEEEoverridecommandlockouts
\usepackage{cite}
\usepackage{bm}
\usepackage{url}
\usepackage{comment}
\usepackage{amsmath,amssymb,amsfonts}
\usepackage{algorithm}
\usepackage{algpseudocode}
\usepackage{graphicx}
\usepackage{multirow}
\usepackage[normalem]{ulem}
\usepackage{longtable}
\usepackage{tabularray}
\usepackage{booktabs}
\usepackage{textcomp}
\usepackage{array}
\newcolumntype{H}{>{\setbox0=\hbox\bgroup}c<{\egroup}@{}}
\usepackage{subcaption}
\usepackage{xcolor}
\newtheorem{definition}{Definition}
\newtheorem{theorem}{Theorem}
\newtheorem{proposition}{Proposition}
\def\BibTeX{{\rm B\kern-.05em{\sc i\kern-.025em b}\kern-.08em
    T\kern-.1667em\lower.7ex\hbox{E}\kern-.125emX}}

\begin{document}

\title{DP-TabICL: In-Context Learning with Differentially Private Tabular Data\thanks{* Equal Contribution}}

\author{\IEEEauthorblockN{Alycia N. Carey*, Karuna Bhaila*, Kennedy Edemacu, Xintao Wu}
\IEEEauthorblockA{\textit{Department of Electrical Engineering and Computer Science} \\
\textit{University of Arkansas}\\
\{ancarey, kbhaila, kedemacu, xintaowu\}@uark.edu}
}

\maketitle

\begin{abstract}
In-context learning (ICL) enables large language models (LLMs) to adapt to new tasks by conditioning on demonstrations of question-answer pairs and it has been shown to have comparable performance to costly model retraining and fine-tuning. Recently, ICL has been extended to allow tabular data to be used as demonstration examples by serializing individual records into natural language formats. However, it has been shown that LLMs can leak information contained in prompts, and since tabular data often contain sensitive information, understanding how to protect the underlying tabular data used in ICL is a critical area of research. This work serves as an initial investigation into how to use differential privacy (DP) -- the long-established gold standard for data privacy and anonymization -- to protect tabular data used in ICL. Specifically, we investigate the application of DP mechanisms for private tabular ICL via data privatization prior to serialization and prompting. We formulate two private ICL frameworks with provable privacy guarantees in both the local (LDP-TabICL) and global (GDP-TabICL) DP scenarios via injecting noise into individual records or group statistics, respectively. We evaluate our DP-based frameworks on eight real-world tabular datasets and across multiple ICL and DP settings. Our evaluations show that DP-based ICL can protect the privacy of the underlying tabular data while achieving comparable performance to non-LLM baselines, especially under high privacy regimes.
\end{abstract}


\section{Introduction}
\label{sec:intro}
Large language models (LLMs) have an impressive ability to perform in-context learning (ICL) \cite{brown2020language}. Conditioned on natural language prompts containing question-answer pairs (called \textit{demonstrations} or \textit{exemplars}), LLMs can execute natural language inference tasks in new domains without having to update the model's parameters while achieving similar downstream performance as full model fine-tuning \cite{duan2023privacy, dong2022survey}. The inference tasks can range from simple fact-retrieval \cite{brown2020language} to complex reasoning and problem-solving \cite{zhou2022teaching} and are adaptable across a wide range of domains. The ease of usage and cost-benefit has motivated several organizations to integrate LLMs into their operations and services to supplement their private data with knowledge from the large corpus of texts that LLMs are trained on. 

Moreover, such organizational data are oftentimes stored in relational databases where data points correspond to individual rows in a table/relation. As one of the most widely used formats of data expression and storage, tabular data has also been used for analytic and predictive tasks in machine learning \cite{borisov2021tabdata}. Recent advances in foundation models and natural language processing (NLP) domains have encouraged researchers to transform or serialize tabular data into natural language formats thus enabling ICL to be performed with tabular data. Authors in \cite{hegselmann2023tabllm} empirically verified nine different methods of serializing tabular data into human-readable strings for use in tabular data classification with LLMs such as manual template-based conversions, table-to-text methods using fine-tuned LLMs, and LLM inference-based methods. Their evaluation shows that LLMs can obtain non-trivial performance in classification tasks for multiple serialized tabular datasets with few-shot fine-tuning as well as zero-shot prompting.   

\begin{figure*}[ht!]
    \centering
    \includegraphics[width=\textwidth]{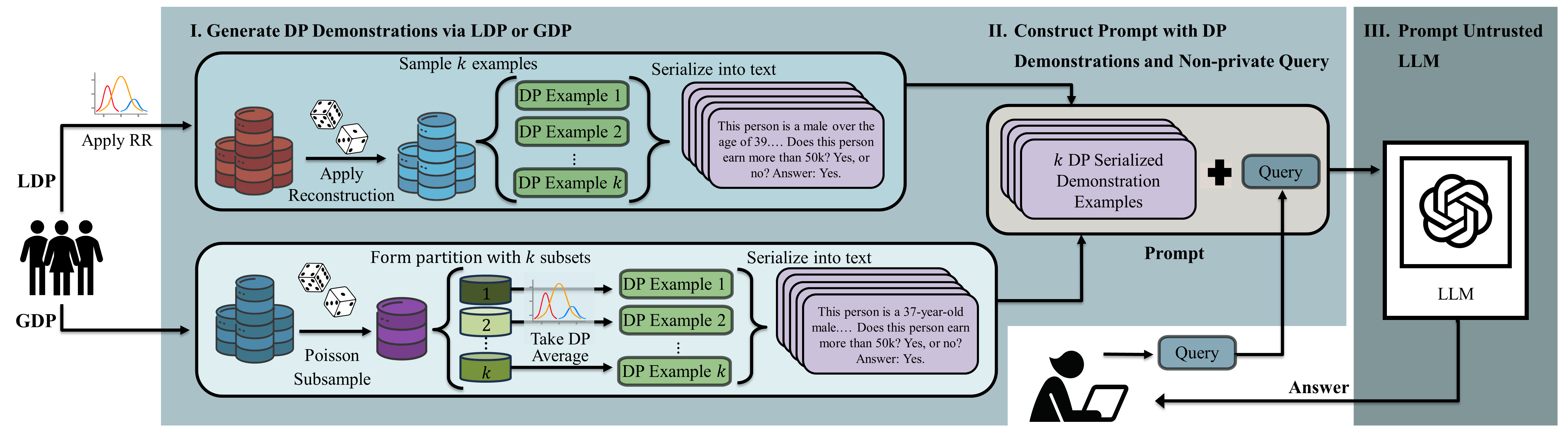}
    \caption{Overview of our two proposed methods methods. In Phase I, DP-protected demonstration examples are generated using either LDP-TabICL (top left) or GDP-TabICL (bottom left). In Phase II, the generated DP-protected demonstration examples are appended with a non-private query from a user to construct the prompt. In Phase III, the constructed prompt is sent to an untrusted LLM and the answer is returned to the user. More details are presented in Section \ref{sec:method}.} 
    \label{fig:dp-tab-icl}
\end{figure*}

The benefits of integrating tabular data into LLMs, however, do not overshadow the risks associated with the use of data possibly containing private and sensitive information about individuals. Recent research has demonstrated that LLMs can leak information from the large text corpus used to train them and from the smaller pool of domain-specific data used to fine-tune them \cite{duan2023flocks}. Furthermore, in \cite{duan2023privacy}, the authors established that data points used within a prompt are highly vulnerable to membership inference attacks (MIAs) and LLMs leak more information through prompting than fine-tuning. 
Additionally, a malicious adversary may further exploit the memorization ability of LLMs by designing prompts that force the models to disclose information provided in prior demonstrations \cite{tang2023privacy}. 

Motivated by these concerns, several works focus on mitigating the privacy risks posed by LLMs and suggest approaches based on differential privacy (DP) \cite{dwork2014algorithmic}. For instance, \cite{yu2022dpfinetuning} introduced a meta-framework that uses DP stochastic gradient descent (DPSGD) \cite{abadi2016deep} during fine-tuning to protect the privacy of the fine-tuning dataset. 
Further, methods based on DP model ensembling have also been proposed for preserving the privacy of data in prompts \cite{tang2023privacy, duan2023flocks, hong2023dp}. 
These methods, however, assume that LLMs can be trusted with private data and focus on the private release of LLM outputs. 
Moreover, these methods were empirically verified on text data with NLP tasks such as sentiment analysis and text classification, and may not necessarily adapt well to transformed tabular data that are comparatively more concise.

In this work, we investigate how well ICL performs when serialized DP-protected tabular data are used as the demonstration examples. We formulate two DP-based frameworks -- Local Differentially Private Tabular-based In-Context Learning (LDP-TabICL) and Global Differentially Private Tabular-based In-Context Learning (GDP-TabICL) -- to generate demonstration examples for ICL that protect the privacy of the underlying tabular dataset. The general idea of our frameworks are presented in Fig. \ref{fig:dp-tab-icl}. LDP-TabICL (top left in Fig. \ref{fig:dp-tab-icl}) utilizes local differential privacy (LDP) to protect the demonstration examples. In LDP, only the users themselves know the true values of their data as they report noisy variants to data aggregator. For LDP-TabICL, we adopt randomized response (RR) \cite{warner1965randomized} to obfuscate user data before collection due to its simplicity, but any LDP frequency oracle \cite{wang2020ldp} may be implemented for privacy-compliant reporting. Since true user data is never disclosed to the server, serialization and prompting can be performed directly with any of the collected perturbed records. However, directly using the noisy records may adversely affect the inference capability of LLMs due to a lack of coherence between the perturbed features and labels. Therefore, we incorporate frequency estimation techniques \cite{wang2016using} and reconstruct an approximation of the true feature and label distributions before sampling data for serialization and prompting. Our second framework, GDP-TabICL (bottom left in Fig. \ref{fig:dp-tab-icl}), utilizes global differential privacy (GDP) to protect the demonstration examples. In GDP, we assume the existence of a trusted curator who collects the users' original (non-perturbed) data and communicates it to the LLM in a private manner. To create the privatized demonstration examples, GDP-TabICL relies on both Poisson subsampling for privacy amplification and the Laplace mechanism \cite{dwork2014algorithmic} to craft differentially private aggregates that represent the underlying data distribution. To generate the prompts from the permuted individual records (LDP-TabICL) or group statistics (GDP-TabICL), we rely on a manual serialization method (see Section \ref{app:examples} in Appendix). \textit{We note that the focus of this work is on evaluating how ICL performs with DP-protected tabular data, so we use the same serialization method for all evaluations and do not perform any prompt tuning. }

We summarize our main contributions as follows:
\begin{itemize}
    \item We propose LDP-TabICL for generating demonstration examples that have formal local DP guarantees for use in tabular data classification via ICL.
    \item We additionally propose GDP-TabICL for generating demonstration examples that have formal global DP guarantees for use in tabular data classification via ICL.
    \item We empirically verify the performance of LDP-TabICL and GDP-TabICL using two open-source LLMs, Llama-2-13B and Llama-2-7B \cite{touvron2023llama}, on eight tabular datasets, and across several ICL and DP settings. We additionally analyze the settings in which LDP-TabICL might be preferable over GDP-TabICL and vice versa.
\end{itemize}

The remainder of the work is as follows. In Section \ref{sec:rel-work} we begin by detailing the current research landscape as it relates to tabular LLMs and differentially private in-context learning. Then, in Section \ref{sec:prelim} we introduce the required backgrounds on in-context learning and differential privacy to successfully craft our frameworks of LDP-TabICL and GDP-TabICL in Section \ref{sec:method}. After detailing our methods and proving their DP-guarantees in Section \ref{sec:method}, we evaluate the ability of LDP-TabICL and GDP-TabICL to produce accurate classifications on tabular data via ICL in Section \ref{sec:eval}. Finally, we offer our concluding remarks in Section \ref{sec:conclusion}.


\section{Related Work}
\label{sec:rel-work}
\paragraph*{\textbf{Tabular LLMs}}
While the majority of works concerning LLMs focus on natural language tasks, there has been a recent push to understand how LLMs can be leveraged to perform tasks over other data modalities such as tabular data. LLMs were shown to be successful at performing semantic parsing \cite{yin2020tabert}, entity matching \cite{li2020deep}, data enrichment \cite{harari2022few}, feature embedding \cite{bertsimas2022tabtext}, data cleaning \cite{min2022rethinking}, and generation \cite{borisov2022language} for tabular data. Additionally, several works explored the ability of LLMs to perform classification tasks on tabular data that have been transformed into natural language formats \cite{hegselmann2023tabllm, dinh2022lift, zhang2023towards, slack2023tablet}. In \cite{dinh2022lift}, the authors proposed Language-Interfaced Fine Tuning (LIFT) which showed that fine-tuning GPT models on multiple tabular datasets leads to classification performance comparable with traditional machine learning models. Further, \cite{hegselmann2023tabllm} showed that using very few-shot fine-tuning offered additional performance increase when tabular data is serialized using simple text templates. Additionally, both \cite{dinh2022lift} and \cite{hegselmann2023tabllm} explored the effects of using different serialization methods to transform the tabular data into natural language. Building off the serialization work of \cite{dinh2022lift} and \cite{hegselmann2023tabllm}, \cite{slack2023tablet} proposed TABLET, a benchmark of 20 tabular datasets that were annotated with instructions varying in phrasing, granularity, and technicality, and showed that using instructions with ICL only (i.e., no fine-tuning) raised the classification ability of LLMs. More recently, \cite{zhang2023towards} proposed TabFM which extends traditional LLMs to be foundation models for tabular data by fine-tuning the LLM on a wide range of tabular datasets to stimulate the acquisition of general knowledge essential for understanding tabular data. We note that while our work also considers the ability of LLMs to perform classification over tabular data via ICL, our work is fundamentally different from the previously proposed works as \textit{we are the first to consider ways to preserve the privacy of the underlying tabular dataset}.

\paragraph*{\textbf{Differentially Private In-Context Learning}}
Due to the wide scale adoption and deployment of LLMs, many works have sought to understand how LLMs can leak private information and have shown that LLMs can memorize data from the underlying training data \cite{carlini2022quantifying, ippolito2022preventing, kharitonov2021bpe, mccoy2023much, wang2024decodingtrust}, from data used in fine-tuning \cite{mireshghallah2022memorization}, and from data used for ICL \cite{priyanshu2023chatbots, duan2023flocks, duan2023privacy, wang2024decodingtrust}. As a result of these findings, researchers have sought to provide data privacy guarantees for all stages of the LLM process, most commonly using DP \cite{dwork2014algorithmic}. For both the training and fine-tuning phase, DPSGD \cite{abadi2016deep} has been the most popular approach \cite{anil2021large, hoory2021learning, li2021large, yu2022dpfinetuning, bu2022differentially} despite it requiring extensive hyperparamter tuning and vast computational resources when applied to LLM models \cite{li2021large, duan2023flocks}. 
For ICL, \cite{duan2023privacy} first studied the privacy leakage through the lens of membership inference attacks (MIAs) and proposed that performing ensembling over different versions of a prompted model can reduce the risk of MIA tasks being successful. A few works \cite{duan2023flocks, wu2023privacy, tang2023privacy}, which were built upon ensembling \cite{papernot2018scalable},  have been proposed to preserve the differential privacy of the underlying prompt data. \cite{tang2023privacy} developed an algorithm that generates differentially private synthetic few-shot demonstrations from the original private dataset and uses the generated samples as demonstrations in ICL inference. The approach leverages the capabilities of multiple local LLMs, feeds each LLM with a disjoint subset of private data, and then privately aggregates to generate private and similar examples. Similarly, \cite{hong2023dp} developed a framework called differentially-private offset prompt tuning (DP-OPT) that combines prompt ensembling, DP, and a local LLM to generate private demonstration examples using forward-backward templates \cite{sordoni2023deep}.  \cite{duan2023flocks} proposed two methods, one based on ensembling (PromptPATE) and one based on DPSGD (PromptDPSGD), for generating hard and soft prompts, respectively. The PromptPATE assumes the existence of an unlabeled public dataset and employs an ensemble of teachers to privately label the public dataset via ICL using the private data. The labeled public dataset is then leveraged as demonstration examples to perform ICL. All these works \cite{duan2023flocks, wu2023privacy, tang2023privacy} assume the use of local LLMs to produce differentially private prompts and incur higher computational cost than our method that uses reconstruction from perturbed data or perturbed statistics to generate DP demonstration examples.   
Recently, the authors of \cite{wu2023privacy} proposed a differentially private in-context learning (DP-ICL) paradigm where the sensitive dataset used for demonstrations is stored in the LLM site rather than the user site studied in our work. Their developed approach performs parallel inference over an ensemble of LLM responses based on disjoint demonstration example subsets and returns the differentially private aggregate. Moreover, all of these works focus on text-based data while \textit{our work is the first to consider protecting tabular data used for ICL}. In addition to considering protection for one of the most widely used forms of data expression in real-world settings, \textit{this allows us to perform LDP/GDP directly on the data} before it is serialized into text form rather than applying it on the output of an LLM ensemble, which allows us to save computational resources and time as we only have to prompt the LLM once per query.

\section{Preliminaries}
\label{sec:prelim}
In this section, we introduce the required background on ICL and DP required to formulate our two DP-TabICL methods. Table \ref{tab:Notation} lists the commonly used notations in this work. Let $\mathcal{X}, \mathcal{Y}$ represent the feature and label domains, respectively, where $\mathcal{X} = \{x_1, \dots, x_F\}$ and $\mathcal{Y} = \{0, 1\}$ as we only consider binary classification. Let $\mathcal{D} = (\mathcal{X}, \mathcal{Y})$ be our dataset of paired features and labels and let it be of size $N$. Similarly, $\hat{\mathcal{D}}= (\hat{\mathcal{X}}, \hat{\mathcal{Y}})$ is our DP-protected dataset, and $\tilde{\mathcal{D}} = (\tilde{\mathcal{X}}, \tilde{\mathcal{Y}})$ is our reconstructed dataset for the LDP-TabICL setting. We let $\mathcal{S}$ be the random subset selected from $\mathcal{D}$ used in GDP-TabICL and we let it be of size $n$. $\mathcal{Q}$ represents a query, $\mathcal{A}$ represents an answer, and $\mathcal{E} = (\mathcal{Q}, \mathcal{A})$ is a demonstration example of a paired query and answer. Here we let $k$ equal the number of shots (demonstration examples) used per prompt. We use the standard notation of $\epsilon, \delta$ to denote our privacy budget and error probability for DP and let $\mathbf{P}$ represent a distortion matrix used in randomized response (RR) for LDP-TabICL. To denote random sampling or selecting random subsets, we use $\in_R$ and $\subset_R$. All lower case letters $x$ denote a  scalar, bold lower case letters $\bm{x}$ denote a vector, and upper case bold variables $\mathbf{X}$ denote matrices.

\begin{table}[t!]
    \centering
    \caption{Notation}
    \label{tab:Notation}
    \begin{tabular}{cl}\toprule
        \textbf{Symbol} & \multicolumn{1}{c}{\textbf{Meaning}} \\ \midrule
        $\mathcal{X}, \mathcal{Y}$ & Feature/label domain\\
        $\mathcal{D} = (\mathcal{X}, \mathcal{Y})$ & Original dataset \\
        $\hat{\mathcal{D}}= (\hat{\mathcal{X}}, \hat{\mathcal{Y}})$ & DP-protected dataset \\
        $\tilde{\mathcal{D}}= (\tilde{\mathcal{X}}, \tilde{\mathcal{Y}})$& Reconstructed dataset for LDP-TabICL\\
        $\in_R$, $\subset_R$ & Random sample/subset\\
        $\mathcal{S}\subset_R\mathcal{D}$ & Random subset of dataset for GDP-TabICL\\
        $\mathcal{Q}, \mathcal{A}$ & Query, answer \\
        $\mathcal{E} = (\mathcal{Q}, \mathcal{A})$ & Demonstration example\\ 
        $k$ & Number of $\mathcal{E}$ for ICL\\
        $\epsilon$, $\delta$ & Privacy budget/error probability for DP \\
        $\mathbf{P}$ & Distortion matrix for RR\\
        $x, \bm{x}, \bm{X}$ & Scalar, vector, matrix\\
        \bottomrule
    \end{tabular}
\end{table}

\subsection{In-Context Learning}
ICL enables large language models to adapt to domain-specific information without requiring the LLM to be fine-tuned or retrained from scratch \cite{wu2023privacy, duan2023privacy}. Specifically, by appending the query $\mathcal{Q}$ to a series of $k$ question-answer demonstration examples $\bm{\mathcal{E}} = [\mathcal{E}_1 = (\mathcal{Q}_1, \mathcal{A}_1), \mathcal{E}_2 = (\mathcal{Q}_2, \mathcal{A}_2), \dots, \mathcal{E}_k=(\mathcal{Q}_k, \mathcal{A}_k)]$, and styling it to fit an appropriate natural language format, we aid the LLM in identifying the mapping between $\mathcal{Q}$ and $\mathcal{A}$. In turn, this enhances the performance of the next token prediction task $\arg\max_{\mathcal{A}}\textbf{LLM}(\mathcal{A}\mid\bm{\mathcal{E}}+\mathcal{Q})$ to generate an answer $\mathcal{A}$ for the wanted query $\mathcal{Q}$. In this work, since we consider tabular data, we first perform serialization to transform the tabular data into a comprehensible sentence before it is used as a query or demonstration example (see Section \ref{sec:serial}).

\subsection{Differential Privacy}
Recent work has shown that ICL is not immune to privacy attacks \cite{duan2023privacy}. \textit{In this work, we aim to increase the privacy of the demonstration examples $\mathcal{E}$ only, and consider the query $\mathcal{Q}$ to be non-sensitive.} To do so, we use DP, which is the current gold standard for data anonymization and protection. DP is formally defined as follows:

\begin{definition}[$(\epsilon,\delta)$-DP \cite{dwork2014algorithmic}]
\label{def:dp}
A randomized algorithm $\mathcal{M}$ with domain $\mathbb{N}^{|\mathcal{X}|}$ is $(\epsilon, \delta)$-DP if for all $\mathcal{O}\subseteq$ Range$(\mathcal{M})$ and for all datasets $\mathcal{D}, \mathcal{D}'\in \mathbb{N}^{|\mathcal{X}|}$ s.t. $||\mathcal{D}-\mathcal{D}'||_1\leq1$:
\begin{equation*}
    \mathbb{P}[\mathcal{M}(\mathcal{D})\in\mathcal{O}] \leq e^{\epsilon}\cdot\mathbb{P}[\mathcal{M}(\mathcal{D}') \in\mathcal{O}]+\delta
\end{equation*}
where $\epsilon$ is the desired privacy budget and $\delta$ is a small error probability.
\end{definition}
In other words, DP ensures that the result of a query over a dataset $\mathcal{D}$ is not overly dependent on a single record. Note, in this work we assume that $\delta=0$ and therefore our discussions will be in terms of ($\epsilon$,0)-DP, which is commonly denoted simply as $\epsilon$-DP. There are two main variants of DP: local and global DP. 

\subsubsection{Local Differential Privacy} 
\label{sec:ldp}
LDP eliminates the necessity of a trusted data curator by requiring end users to perturb their data before sending it to the server. In this work, we use the popular LDP method of randomized response (RR) to protect the original data \cite{warner1965randomized, wang2016using}. Let $u$ be a private variable that can take one of $U$ values. We can formalize the RR process as a $U\times U$ distortion matrix $\mathbf{P}=(p_{uu'})_{U\times U}$ where $p_{uu'}=\mathbb{P}[u'\mid u]\in(0,1)$ denotes the probability that the output of the RR process is $u'\in\{1,\dots,U\}$ when the real attribute value is $u\in\{1,\dots,U\}$. Note that $\mathbf{P}$ can be set to achieve optimal utility and $\epsilon$-DP by setting $p_{uu'}=\frac{e^\epsilon}{U-1+e^\epsilon}$ when $u = u'$ and $p_{uu'}=\frac{1}{U-1+e^\epsilon}$ otherwise \cite{wang2016using}. In this work, we allow each feature $f = \{1, \dots, F\}$ and label $y$ to have its own distortion matrix $\mathbf{P}_i$ for $i \in \{1,\dots,F,y\}$ set by each feature/label's allocated privacy budget $\epsilon_i$ (where $\epsilon=\sum\epsilon_i$) and domain $d_i$ (where $d_i=U$).

If we have a dataset that was collected under RR, then using the distortion matrices $\mathbf{P_i}$ used during data collection, we can estimate the true population distribution from the noisy collected data. Let $\bm{\pi}=[\pi_{1, \dots, 1}, \dots, \pi_{d_1, \dots, d_F, d_y}]$ be the true proportion of the values in the original population and let $\bm{\lambda} = [\lambda_{1, \dots, 1}, \dots, \lambda_{d_1, \dots, d_F, d_y}]$ be the observed proportion of the values in the collected noisy dataset. Using the relationship $\bm{\pi}\approx\bm{\mathcal{P}}^{-1}\bm{\lambda}$, where $\bm{\mathcal{P}}^{-1} = \mathbf{P}_1^{-1} \otimes\dots\otimes\mathbf{P}_F^{-1}\otimes\mathbf{P}_y^{-1}$, we can estimate the true underlying population $\bm{\pi}$ based on the observed values in the collected noisy dataset $\bm{\lambda}$. Here, $\otimes$ denotes the Kronecker product of two matrices.

\subsubsection{Global Differential Privacy} In GDP, we assume a trusted data curator collects the sensitive data and ensures that any query over the data (e.g., count, average, \dots) is made differentially private by adding noise proportional to the sensitivity of the query. In Section \ref{sec:gdp-tabicl}, we use the Laplace mechanism to generate differentially private averages. 
\begin{definition}[Laplace Mechanism \cite{dwork2014algorithmic}]
\label{def:laplace}
Given any function $f:\mathbb{N}^{|\mathcal{X}|}\to\mathbb{R}^c$, the Laplace mechanism is defined as:
\begin{equation}
    \mathcal{M}_L(x, f(\cdot), \epsilon) = f(x) + (v_1, \dots, v_c)
\end{equation}
where $v_i$, $i \in \{1, \dots, c\}$ are i.i.d. random variables drawn from Lap($\Delta f/\epsilon)$ and $\Delta f$ is the sensitivity of the function $f$.
\end{definition}

\subsubsection{Properties of Differential Privacy}
There are several properties of DP that we take advantage of when constructing our DP-TabICL algorithms in Section \ref{sec:method}. 

\textbf{Sequential composition} bounds the total privacy cost $\epsilon$ of performing multiple DP queries over the same input dataset. 
\begin{theorem}[Sequential Composition \cite{dwork2014algorithmic}]
\label{thm:seq}
Let $\mathcal{M}_1:\mathbb{N}^{|\mathcal{X}|}\to\mathcal{R}_1$ be an $\epsilon_1$-DP algorithm, and let $\mathcal{M}_2:\mathbb{N}^{|\mathcal{X}|}\to\mathcal{R}_2$ be an $\epsilon_2$-DP algorithm. Then their combination, defined to be $\mathcal{M}_{1,2}:\mathbb{N}^{|\mathcal{X}|}\to\mathcal{R}_1\times\mathcal{R}_2$ by the mapping $\mathcal{M}_{1,2}(\mathcal{D})=(\mathcal{M}_1(\mathcal{D}), \mathcal{M}_2(\mathcal{D})) $ is $\epsilon_1+\epsilon_2$-DP.
\end{theorem}

\textbf{Parallel composition} is another way to bound the total privacy cost of multiple data releases. In parallel composition, the dataset is split into disjoint subsets and the DP mechanism is ran on each subset independently. 
\begin{theorem}[Parallel Composition] 
\label{def:para}
Let $\mathcal{M}:\mathbb{N}^{|\mathcal{X}|}\to\mathcal{R}$ be an $\epsilon$-DP algorithm. Let $\mathcal{D}_1 \cup \mathcal{D}_2=\mathcal{D}$ be disjoint subsets. Then, the mechanism that releases all of the results $\mathcal{M}(\mathcal{D}_1, \mathcal{D}_2)=(\mathcal{M}(\mathcal{D}_1), \mathcal{M}(\mathcal{D}_2))$ is $\epsilon$-DP.
\end{theorem}

\textbf{Post-processing} states that it is impossible to reverse the privacy protection provided by DP by post-processing the data in any manner despite the availability of auxiliary information.
\begin{proposition}[Post-processing \cite{dwork2014algorithmic}]
\label{prop:post-proc}
Let $\mathcal{M}:\mathbb{N}^{|\mathcal{X}|}\to\mathcal{R}$ be a randomized algorithm that is $\epsilon$-differentially private. Let $f:\mathcal{R}\to\mathcal{R}'$ be an arbitrary randomized mapping. Then $f\circ\mathcal{M}:\mathbb{N}^{|\mathcal{X}|}\to\mathcal{R}'$ is $\epsilon$-differentially private.
\end{proposition}

\textbf{Privacy amplification by subsampling} says that the privacy guarantees of a differentially private mechanism can be amplified by applying it only to a small random subsample of records from a given dataset \cite{balle2018privacy}. 
\begin{definition}[Differential Privacy under Sampling \cite{li2012sampling}]
\label{def:dps}
An algorithm $\mathcal{M}$ gives $(\beta, \epsilon, \delta)$-DPS (DP under sampling) if and only if $\beta > \delta$ and the algorithm $\mathcal{M}^\beta$ gives $(\epsilon, \delta)$-DP, where $\mathcal{M}^\beta$ denotes the algorithm to first sample with probability $\beta$ (include each tuple in the input dataset with probability $\beta$), and then apply $\mathcal{M}$
to the sampled dataset. 
\end{definition}
\begin{theorem}[Amplification Effect of Sampling \cite{li2012sampling}]
\label{thm:amplification}
    Any algorithm that satisfies $(\beta_1, \epsilon_1, \delta_1)$-DPS also satisfies $(\beta_2, \epsilon_2, \delta_2)$-DPS for any $\beta_2<\beta_1$ where $\epsilon_2=\ln\left(1+\left(\frac{\beta_2}{\beta_1}(e^{\epsilon_1}-1)\right)\right)$, and $\delta=\frac{\beta_2}{\beta_1}\delta_1$.
\end{theorem}

\section{Proposed Method} 
\label{sec:method}
To preserve the privacy of tabular data used for ICL, we propose two DP-based strategies. The first, Local Differentially Private Tabular-based In-Context Learning (LDP-TabICL), relies on the local DP method of randomized response while the second, Global Differentially Private Tabular-based In-Context Learning (GDP-TabICL), uses global DP via the Laplace method. An overview of our (L/G)DP-TabICL processes can be seen in Fig. \ref{fig:dp-tab-icl}. 

\subsection{LDP-TabICL}
\label{sec:ldp-tabicl}
LDP-TabICL utilizes randomized response (see Section \ref{sec:ldp}) to protect the underlying tabular data being used to generate demonstration examples. In this setting, we assume that the data was originally collected using the randomized response process and that \textit{we do not have access to the original data $\mathcal{D}$.} However, we assume that the distortion matrices $\mathbb{P} = [\mathbf{P}_1, \dots, \mathbf{P}_F, \mathbf{P}_y]$ used during data collection are known.  Using the known distortion matrices, before serializing the tabular data into demonstration examples, we perform reconstruction over the perturbed data to create a dataset that is more representative of the original population. 

\begin{algorithm}[h]
\caption{LDP-TabICL}
\label{alg:ldp-tabicl}
\begin{algorithmic}[1]
    \renewcommand{\algorithmicrequire}{\textbf{Input:}}
    \Require Tabular dataset: $\mathcal{D}$, distortion matrices: $\mathbb{P} = [\mathbf{P}_1, \dots, \mathbf{P}_F,\mathbf{P}_y]$, num. shots: $k$, serialization function: \textbf{SF}, privacy budget $\epsilon$
    \renewcommand{\algorithmicrequire}{\textbf{Output:}}
    \Require Privacy-preserving prompts
    \State $\hat{\mathcal{D}} \gets$ Collect $\mathcal{D}$ according to $\epsilon$ randomized response
    \State $\tilde{\mathcal{D}} \gets$ Perform reconstruction on $\hat{\mathcal{D}}$ according to $\mathbb{P}$
    \State $\tilde{\bm{z}} \gets \{(\tilde{\bm{x}}, \tilde{y})\}^k\in_R \tilde{\mathcal{D}}$
    \State $\bm{\mathcal{E}}$ = \textbf{SF}$(\tilde{\bm{z}})$

    \State \textbf{Return:} $\bm{\mathcal{E}}$
\end{algorithmic}
\end{algorithm}
The pseudo-code of our LDP-TabICL method can be seen in Alg. \ref{alg:ldp-tabicl} and a visual depiction can be seen in Fig. \ref{fig:dp-tab-icl}. 
First, in line 1 we collect the perturbed data records that were locally randomized using the preset distortion matrices $\mathbb{P} = [\textbf{P}_1, \dots, \textbf{P}_F,\textbf{P}_y]$. 
In this manner, we do not have access to the original data $\mathcal{D}$, but rather only to the DP-protected data $\hat{\mathcal{D}}$. Next, in line 2 we use the known distortion matrices on the randomized dataset $\hat{\mathcal{D}}$ to reconstruct an approximation $\tilde{\mathcal{D}}$ to the original, un-randomized dataset $\mathcal{D}$ (see Section \ref{sec:ldp}). Then, based on the number of desired shots $k$, we randomly sample $k$ examples from $\tilde{\mathcal{D}}$ (line 3). Finally, we serialize the selected examples $\tilde{\bm{z}}$ into natural language via our serialization function (line 4), and return the generated set of demonstration examples $\bm{\mathcal{E}}$ to be used in conjunction with a query $\mathcal{Q}$ to prompt an LLM. In Theorem \ref{thm:ldp} we show that LDP-TabICL satisfies $\epsilon$-DP.

\begin{theorem}
\label{thm:ldp}
    LDP-TabICL satisfies $\epsilon$-DP. 
\end{theorem}
\begin{IEEEproof}
Assume that each feature $f = \{1, \dots, F\}$ and the label $y$ are each allotted $\epsilon_i$, $i \in \{1, \dots, F, y\}$, as its privacy parameter where $\epsilon = \sum \epsilon_i$. Then, via sequential composition (Thm. \ref{thm:seq}), the total privacy budget is $\epsilon$, and assuming $\delta=0$, LDP-TabICL satisfies $\epsilon$-DP.
\end{IEEEproof}

\subsection{GDP-TabICL}
\label{sec:gdp-tabicl}
Whereas LDP-TabICL uses local DP to provide protection to the tabular dataset, GDP-TabICL relies on global DP. Here, we assume that we have access to the original dataset $\mathcal{D}$ and that it has not undergone randomization or reconstruction. We detail the pseudo-code of our GDP-TabICL method in Alg. \ref{alg:gdp-tabicl} and a visual depiction can be seen in Fig. \ref{fig:dp-tab-icl}.

\begin{algorithm}
\caption{GDP-TabICL}\label{alg:gdp-tabicl}
\begin{algorithmic}[1]
    \renewcommand{\algorithmicrequire}{\textbf{Input:}}
    \Require Tabular dataset: $\mathcal{D}$, size of $\mathcal{D}$: $N$, size of subsample: $n$, num. shots: $k$, serialization function: \textbf{SF}, num. features: $F$, privacy budget per feature/label $\bm{\epsilon} = [\epsilon_1, \dots, \epsilon_F, \epsilon_y]$
    \renewcommand{\algorithmicrequire}{\textbf{Output:}}
    \Require Privacy-preserving prompts
    \State $\bm{\mathcal{E}} = [\;]$
    \State $\mathcal{S} \gets \{(\bm{x}, y)\}^n\in_R \mathcal{D}$
    \State $S_1 \cup \dots \cup S_k = \mathcal{S}$ s.t. $S_1 \cap \dots \cap S_k = \emptyset$
    \For{$i = 1 \dots k$}
        \For{$j = 1, \dots, F$}
            \State $\hat{x}_j \gets$ DP\_Avg($[x_j \in S_i]$, $\epsilon_j$)
        \EndFor
        \State $\hat{\bm{x}} = [\hat{x}_1, \dots, \hat{x}_F]$
        \State $\hat{y} \gets$ DP\_Avg($[y \in S_i]$, $\epsilon_y$)
        \State $\bm{\mathcal{E}}$.append(\textbf{SF}($\hat{\bm{x}}, \hat{y}$))
    \EndFor
    \State \textbf{Return:} $\bm{\mathcal{E}}$
\end{algorithmic}
\end{algorithm}

In Alg. \ref{alg:gdp-tabicl} we begin by taking a random sample $\mathcal{S}$ of size $n$ from the original dataset $\mathcal{D}$  (line 2). Then, in line 3 we create $k$ subsets that form a partition of $\mathcal{S}$ where $k$ is the number of demonstration examples we want to include in the prompt. For each created subset $S_i$ in the partition, we go through each of the features $x_f$ and compute the differentially private average $\hat{x}_f$ (line 6). After computing the differentially private average for each feature, we take the differentially private average $\hat{y}$ of the labels $y$ (line 9). We can then construct a demonstration example from the calculated differentially private averages by feeding $\hat{\bm{x}} = \{\hat{x}_1, \dots, \hat{x}_F\}$ and $\hat{y}$ to the serialization function in line 10. After crafting the differentially private demonstration example $\mathcal{E}_i$ for each subset $S_i$, we return the demonstration examples $\bm{\mathcal{E}} = [\mathcal{E}_1, \dots, \mathcal{E}_k]$ so they can be used in conjunction with a query $\mathcal{Q}$ to prompt an LLM. We discuss how we specifically create the partition of $\mathcal{S}$ and compute the differentially private average of each subset in Section \ref{sec:exp-gdptabicl}. In Theorem \ref{thm:gdp-tabicl} we show that due to subsampling GDP-TabICL achieves $\textrm{ln}\left(1 + (\frac{n}{N}(e^\epsilon - 1)\right)$-DP.

\begin{theorem}
\label{thm:gdp-tabicl}
    GDP-TabICL satisfies $\textrm{ln}\left(1 + (\frac{n}{N}(e^\epsilon - 1)\right)$-DP. 
\end{theorem}
\begin{IEEEproof}
Let $\mathcal{M}$ be a differentially private averaging function that satisfies $\epsilon$-DP (i.e., $\delta=0$). First, consider GDP-TabICL where sampling $\mathcal{S} \in_R \mathcal{D}$ is \textit{not} performed and creating the subsets $S_1,\dots, S_k$ to form a partition of $\mathcal{S}$ is performed directly on $\mathcal{D}$. Then, the computation of a differentially private-averaged prompt via $\mathcal{M}$ for a single subset $S_i$ is $\epsilon$-DP. However, since all subsets $S_1,\dots,S_k$ in the partition are disjoint, by parallel processing (Def. \ref{def:para}) the total privacy budget to compute all $k$ DP-averaged prompts is $\epsilon$-DP. Now consider the case where sampling $\mathcal{S}\in_R\mathcal{D}$ is performed. Due to privacy amplification by subsampling (Thm. \ref{thm:amplification}), the overall privacy budget is $\textrm{ln}\left(1 + (\frac{n}{N}(e^\epsilon - 1)\right)$-DP.
\end{IEEEproof}

\subsection{Serialization}
\label{sec:serial}
As the purpose of this work is to understand how DP can be incorporated into querying LLMs with tabular data, we choose a fairly simple serialization method to transform the protected tabular data into natural language statements. Specifically, for each dataset, we craft a custom text template that will be filled in with the protected tabular data. For example, the template we constructed for the blood dataset in the GDP case is as follows: 

\textit{A blood donor at a Blood Transfusion Service Center is described as follows: The donor has donated blood [frequency] times. They have donated a total of [monetray] c.c. of blood. They last donated blood [recency] months ago. Their first blood donation was [time] months ago. Did the donor donate blood in March 2007? Yes or No? Answer: [label].}

Here, \textit{frequency, monetray, recency} and \textit{label} were filled in according to the DP tabular data whether it was produced using the reconstructed randomized data (LDP-TabICL) or from differentially private aggregates (GDP-TabICL). We provide examples of the constructed prompts for all datasets in both LDP-TabICL and GDP-TabICL cases in the Appendix \ref{app:examples}.

\section{Experimental Evaluation}
\label{sec:eval}
In this section, we analyze the overall effect of using LDP-TabICL and GDP-TabICL to preserve the privacy of tabular-based demonstration examples for ICL. We begin by describing our experimental settings and then discuss specific empirical results. We note that we run each experiment five times and report the average and standard deviation for all reported values. We vary the privacy budget, $\epsilon$, among $\{1, 5, 10, 25, 50\}$ and set $\delta=0$ for simplicity in our experiments and evaluate the frameworks for few-shot classification across a varied number of demonstrations, $k \in \{1, 2, 4, 8\}$. We additionally note that all experiments were run on a Tesla V100 GPU (32GB RAM) and a Xeon 6258R 2.7GHz CPU. 
We report the total time for all experimental settings as well as their carbon emissions in Appendix \ref{app:comp-cost}.

\subsection{Datasets}
We consider eight common tabular datasets in our evaluation of LDP-TabICL and GDP-TabICL: adult (48842 rows, 12 feats.), bank (45211 rows,  16 feats.), blood (748 rows, 4 feats.), calhousing (20640 rows, 8 feats.), car (1728 rows, 6 feats.), diabetes (768 rows, 8 feats.), heart (918, 11 feats.), and jungle (44819 rows, 6 feats.). Out of these datasets four are relatively balanced (calhousing: 50\%/50\%, diabetes: 60\%/40\%, heart: 55\%/45\%, jungle: 51\%/49\%) while the other four are unbalanced (adult: 75\%/25\%, bank: 88\%/12\%, blood: 75\%/25\%, car: 70\%/30\%). We consider only binary classification and therefore change the multi-class datasets (calhousing, car, jungle) to only contain two classes. We split each dataset into 80\% to be used in crafting demonstration examples and 20\% to be used as the unlabeled test query. We note that due to the cost and time required to prompt an LLM, we only test 10\% of the unlabeled test queries for the large datasets (adult, bank, calhousing, and jungle), while we test only 75\% for small datasets (blood, car, diabetes, heart). When testing LDP-TabICL, we assume that all features have been binarized using one-hot encoding for the categorical features and thresholding via the mean for numerical features. We limit the number of features to be $\leq14$ to enable reconstruction to be performed and therefore perform additional feature engineering on the datasets. We refer interested readers to our github\footnote{http://tinyurl.com/remyzjxk} for further detail on our data preprocessing steps. For GDP-TabICL, we use the dataset as given and do not perform binarization or feature selection.

\subsection{LLM and Baseline Models}
In this work, we use a locally deployed pre-trained Llama-2-13B model that has not gone through any additional fine-tuning as the base LLM model to evaluate LDP-TabICL and GDP-TabICL for their inference efficiency with differentially private serialization and prompting. We also conduct experiments using the Llama-2-7B model to analyze the influence, if any, of the size and complexity of the base LLM. For brevity, we only include the 13B results in the main body and leave the results of the 7B model to Section \ref{app:7B} of the Appendix. Additionally, in the LDP scenario, we compare against logistic regression (LR) and Gaussian Na\"ive Bayes (GNB) models as baselines using the full privatized and reconstructed training set as input. For the GDP scenario, we also compare against LR and GNB, but instead of training on LDP-protected and reconstructed data, the training routine itself is modified to achieve global DP. Specifically, we use the GDP implementations of LR and GNB from the Diffprivlib library \cite{diffprivlib}. Note that we train all baseline models with the full training set and compare the results against $k$-shot classification using DP-compliant prompts generated with the proposed frameworks. We further point out that when $\epsilon = \infty$, we still employ LDP and GDP data collection procedures. This means that, for LDP, data is reported with a 0\% probability of perturbation, and for GDP, we obtain the subset aggregates without adding any noise.


\subsection{Serialization}
As mentioned in Section \ref{sec:serial}, we craft custom templates for each dataset into which the DP-protected tabular data can be plugged. We note that in the GDP-TabICL setting, we perform additional post-processing on the differentially private averages to keep the values within a feasible region. For example, since a person cannot be negative years old (which could be generated via DP averaging), we lower bound the DP age to 0. We list examples of serialization for GDP-TabICL where $k=2$ and $\epsilon=5$ in Table \ref{tab:prompt-examples} in Appendix \ref{app:gdp-examples}.

\subsection{LDP-TabICL}
To perform LDP-TabICL according to Alg. \ref{alg:ldp-tabicl}, we first perturb the original dataset $\mathcal{D}$ using $\epsilon$ randomized response to simulate the collection of already randomized data. We note that after perturbing the original dataset, we do not access it again and carry out the remaining steps using $\hat{\mathcal{D}}$. Using the perturbed dataset, we first perform reconstruction\footnote{Note: For efficiency, this step can be done once for all queries rather than once per each query. In our experimentation, we only perform reconstruction once for all queries.} to construct $\tilde{\mathcal{D}}$ and then sample $k$ rows to serialize into demonstration examples. \textit{We note that for each test query $\mathcal{Q}$ we generate a new set of demonstration examples $\bm{\mathcal{E}}$ by sampling from $\tilde{\mathcal{D}}$}. This, however, does not cause additional privacy expenditure as the randomized response and reconstruction process has already been performed over the entire dataset $\mathcal{D}$. \

\subsection{GDP-TabICL}
\label{sec:exp-gdptabicl}
According to Alg. \ref{alg:gdp-tabicl}, we perform the following:

\paragraph*{\textbf{Selecting} $\bm{\mathcal{S}}$}
To select the subset $\mathcal{S}\subset\mathcal{D}$ we use Poisson sampling with rate $\lambda = \frac{n}{N}$. The resulting amplification in privacy can be seen in Table \ref{tab:e-vs-e'} in Appendix \ref{app:amp-by-sub}. 

\paragraph*{\textbf{Dividing} $\bm{\mathcal{S}}$ \textbf{into} $\bm{k}$ \textbf{Subsets}}
\label{sec:div-subset}
There are many different methods to divide $\mathcal{S}$ into $k$ different subsets, such as: using GROUP BY queries, dividing the set uniformly at random, and using Poisson sampling. In this work, we utilize the GROUP BY process to attempt creating representative demonstration examples. Specifically, we divide $\mathcal{S}$ into $k$ subsets by grouping based on features or the label that have possible attribute values that sum to $k$. For example, if $k=2$, we can GROUP BY on the label $y$ to create the two disjoint subsets. 
We discuss specifics of how each dataset was grouped in Appendix \ref{app:group-by}.

\paragraph*{\textbf{DP Average}}
\label{sec:dp-avg}
The procedure to calculate the DP average in GDP-TabICL depends on whether a feature is categorical or numerical. We assume that the total privacy budget $\epsilon$ is split equally between all features $f=1,\dots,F$ and the label $y$ meaning that each feature and the label are allotted $\xi = \frac{\epsilon}{F+1}$ as their privacy budget. \textit{We note that we use the same demonstration examples $\mathcal{E}$ for each test query $\mathcal{Q}$ to avoid spending additional privacy budget.} However, each of the five trials gets a full $\epsilon$ privacy budget.

\paragraph{Categorical}
If a feature (or the label) is categorical, then we can construct a histogram of differentially private counts and select the attribute with the highest value. Due to histograms naturally meeting the criteria for parallel composition (Def. \ref{def:para}), we can allot the entire $\xi$ budget for the features and label to privatize each attribute count. Since the sensitivity of a count query is $\Delta=1$, we add $\textrm{Lap}(\frac{1}{\xi})$ to each attribute count. Then, via post-processing (Prop. \ref{prop:post-proc}), we can select the highest differentially private count without expending any additional privacy.

\paragraph{Numerical}
If a feature is numerical, then to take the differentially private average, we must first take the differentially private count and then the differentially private sum and therefore we divide the allotted privacy budget $\xi$ evenly between the two functions. To take the differentially private count, we add Lap$\left(\frac{1}{\xi/2}\right)$ due to the count query having a sensitivity of $\Delta=1$. Taking the differentially private sum is slightly more complex as the sensitivity is now unbounded. To achieve bounded sensitivity, we employ clipping based on the upper ($\alpha$) and lower ($\gamma$) bounds on what value the feature can take. We assume that $\alpha$ and $\gamma$ are set based on properties of the feature/label that are known without looking into the data itself to simplify calculating the privacy expenditure.\footnote{This assumption may not hold in all settings and additional privacy may be required to compute the differentially private upper and lower bounds.} For example, if the feature contains ages, then we can set $\gamma=0$ and $\alpha=120$. To take the differentially private sum, we add Lap$\left(\frac{\alpha-\gamma}{\xi/2}\right)$ noise to the non-private sum. Then, via post-processing (Prop. \ref{prop:post-proc}), we calculate the average by dividing the differentially private sum by the differentially private count.

\subsection{Analysis}
\subsubsection{LDP-TabICL}
We report the LDP-TabICL experimental results on Llama-2-13B, as well as the results from the two baseline models trained on the reconstructed LDP data, in Table \ref{tab:ldp-tabicl-results-13b}. In general, when using LDP-TabICL to prompt the model with a certain number of demonstration examples $k$, there is no real trend for the accuracy as $\epsilon$ increases. This seems counter-intuitive to the general intuition that data with less noise (larger $\epsilon$) leads to better model performance (as shown in the baseline models' performance across $\epsilon$). However, since we perform reconstruction on the data before sampling the demonstration examples, it makes sense that there is no trend.\footnote{We additionally perform reconstruction on the LDP-protected data before training the baseline models. However, due to training the baselines over the entire training dataset (rather than sampling a few demonstration examples) the effects of noise are more noticeable as the reconstruction only gives a (noisy) estimate of the underlying distribution.} By performing reconstruction we essentially negate the effects of noise by estimating the true underlying data distribution (see Section \ref{sec:ldp}). This means that, in general, for high levels of noise (small $\epsilon$), we are able to craft demonstration examples that when used to prompt an LLM achieve higher levels of accuracy than the baseline models trained under the same $\epsilon$ value. For example, when $k=1$ and $\epsilon=1$ for the adult dataset, LDP-TabICL is able to achieve 78\% accuracy, while the best performing baseline only achieves 38\%. Further, the $p$-value for the one-tailed paired Student t-test for $k=1$ and $\epsilon=1$ across all datasets against the LR baseline is 0.02, and against the GNB baseline is 0.01, meaning that the LLM achieving higher accuracy is not due to chance and is statistically significant. However, we note that at a certain threshold (for example, in these experiments $\epsilon \approx 25$), the baselines do start to perform better than LDP-TabICL as they are able to learn from the entire (mostly noise-free) training dataset while LDP-TabICL is only prompted on a few demonstration examples.

While there is no noticeable trend in the accuracy as $\epsilon$ increase in general, other trends become noticeable when we analyze the performance of LDP-TabICL on the balanced datasets (calhousing, diabetes, heart, jungle) against its performance on the unbalanced datasets (adult, bank, blood, car). Most noticeably, LDP-TabICL performs best on the unbalanced datasets when only a few number of demonstration examples ($k\leq2$) are used, while it performs best on the balanced datasets when more demonstration examples ($k\geq4$) are used. This is because in the small $k$ setting, either majority voting (LLM predicts only one label for all prompts) or random guessing are performed, which leads to the unbalanced datasets naturally achieving higher accuracy. For example, the adult dataset had 4\% true positive (TP) and 74\% true negative (TN) via majority voting (8\% predicted True, 92\% predicted False) when $\epsilon=1$ and $k=1$ which gave the accuracy score of 78\%. Calhousing, on the other hand, had 43\% TP and 10\% TN by random voting (50\% predicted True, 50\% predicted False) when $\epsilon=1$ and $k=1$ which gave the accuracy score of 53\%. In contrast, when more demonstration examples are used (larger $k$ value), the model has more context to learn from and performs better on the balanced datasets. In this case, when $\epsilon=1$ and $k=4$, the adult dataset had 9\% TP and 68\% TN which gave an accuracy score of 77\%, while the calhousing dataset had 14\% TP and 44\% TN giving an accuracy score of 58\%. We additionally note that when $\epsilon=\infty$ (i.e., no privacy protection), the unbalanced datasets have accuracy similar to what is achieved by the baseline models, while the balanced datasets have accuracies that are slightly farther away from the baselines.
 
In addition to the difference in accuracy, the LDP-TabICL results have much higher standard deviations on the unbalanced datasets than the balanced datasets. Further, for the unbalanced datasets in general, when $k\geq4$ LDP-TabICL experiences higher standard deviations than when $k<4$ and $\epsilon>10$. We attribute both of these trends to LLMs being highly sensitive to what it is prompted on. Due to randomly sampling the demonstration examples from the reconstructed LDP-protected dataset, if the dataset is unbalanced there is a higher chance that the demonstration examples selected are not representative of the underlying data distribution. We note, however, that the baselines in this setting generally have higher standard deviations as well meaning that the high standard deviation is not unnatural or a byproduct of using LDP-TabICL. 

\textit{In general, these results show that LLMs can produce similar, or even better, accuracy on LDP-protected tabular data when compared with baseline models trained on the reconstructed LDP-protected data, especially in the settings where there is a high privacy requirement (i.e., $\epsilon$ is small).} We hypothesize that crafting better prompt templates and performing prompt engineering can lead to even better accuracy by the LLM model on the LDP-protected tabular data. We additionally note that while it may not be known a priori how many demonstration examples $k$ should be used to achieve the best accuracy in a specific setting (as noted by the inconsistent trend in accuracy as $k$ increases), since the demonstration examples are only sampled after the data is protected via LDP (and reconstructed), experimenting with different number of demonstration examples to find the number $k$ that gives the best result, or using prompt tuning strategies that select the demonstration examples in an algorithmic way (e.g., using K-means clustering to select examples close to the query) does not incur any additional privacy expenditure.

\begin{table}[t!]
\centering
\caption{LDP-TabICL average accuracy and standard deviation over 5 runs on Llama-2-13B using differing number of splits $k$ and privacy budget value $\epsilon$. LR: logistic regression, GNB: Gaussian Na\"ive Bayes, *: model trained on reconstructed LDP data, \underline{Datset}: unbalanced dataset.}
\label{tab:ldp-tabicl-results-13b}
\resizebox{.95\columnwidth}{!}{%
\begin{tabular}{@{}ccccHccccc@{}}
\toprule
 & \textbf{} & \multicolumn{8}{c}{$\bm{\epsilon}$} \\
\textbf{Dataset} & \textbf{Method} & $\bm{k}$ & \textbf{1} & \textbf{2} & \textbf{5} & \textbf{10} & \textbf{25} & \textbf{50} & $\bm{\infty}$\\ \midrule
\multirow{6}{*}{\underline{Adult}} & LDP-TabICL & 1 & \textbf{0.78\textsubscript{.01}}& 0.73\textsubscript{.10}& 0.74\textsubscript{.06}& \textbf{0.74\textsubscript{.06}}& 0.70\textsubscript{.18}& 0.79\textsubscript{.02}& 0.72\textsubscript{.01}\\
                       & LDP-TabICL & 2 & 0.73\textsubscript{.09}& \textbf{0.78\textsubscript{.01}}&\textbf{0.79\textsubscript{.03}} & 0.69\textsubscript{.24}& 0.78\textsubscript{.03}& 0.77\textsubscript{.02}& 0.72\textsubscript{.01}\\
                       & LDP-TabICL & 4 & 0.77\textsubscript{.04}& 0.77\textsubscript{.05}& 0.61\textsubscript{.21}& 0.72\textsubscript{.17}& 0.75\textsubscript{.07}& 0.78\textsubscript{.01}& 0.73\textsubscript{.01}\\
                       & LDP-TabICL & 8 & 0.68\textsubscript{.15}& 0.73\textsubscript{.12}& 0.66\textsubscript{.11}& 0.70\textsubscript{.21}& 0.76\textsubscript{.04}& 0.76\textsubscript{.01}& 0.75\textsubscript{.02}\\
                       & LR*         & - & 0.35\textsubscript{.20} & 0.59\textsubscript{.12} & 0.54\textsubscript{.16} & 0.70\textsubscript{.06} & \textbf{0.82\textsubscript{.00}} & \textbf{0.83\textsubscript{.00}} & \textbf{0.85\textsubscript{.00}} \\
                       & GNB*         & - & 0.38\textsubscript{.20} & 0.60\textsubscript{.13} & 0.57\textsubscript{.15} & 0.71\textsubscript{.06} & 0.82\textsubscript{.00} & 0.81\textsubscript{.00} & 0.81\textsubscript{.00} \\\midrule
 
\multirow{6}{*}{\underline{Bank}} & LDP-TabICL& 1 & \textbf{0.88\textsubscript{.01}} & \textbf{0.87\textsubscript{.03} }& \textbf{0.88\textsubscript{.02}} & 0.88\textsubscript{.01} & 0.84\textsubscript{.04}& 0.86\textsubscript{.02}& 0.85\textsubscript{.01}\\
                       & LDP-TabICL & 2 & \textbf{0.88\textsubscript{.02} }& 0.77\textsubscript{.24} & \textbf{0.88\textsubscript{.02}} & 0.86\textsubscript{.06} & 0.88\textsubscript{.01} & 0.87\textsubscript{.02} & 0.87\textsubscript{.01} \\
                       & LDP-TabICL & 4 & 0.79\textsubscript{.17}& 0.79\textsubscript{.20}& 0.84\textsubscript{.09}& \textbf{0.89\textsubscript{.01}}& 0.88\textsubscript{.01}& 0.87\textsubscript{.02}& 0.87\textsubscript{.01} \\
                       & LDP-TabICL & 8 & 0.65\textsubscript{.31}& \textbf{0.87\textsubscript{.03}}& 0.80\textsubscript{.14}& 0.77\textsubscript{.18}& 0.86\textsubscript{.05}& 0.82\textsubscript{.11}& 0.85\textsubscript{.01}\\
                      & LR*        & - & 0.57\textsubscript{.37} & 0.45\textsubscript{.35} & 0.67\textsubscript{.14} & 0.88\textsubscript{.01} & \textbf{0.89\textsubscript{.00}} & \textbf{0.89\textsubscript{.00}} & \textbf{0.90\textsubscript{.00}} \\
                      & GNB*         & - & 0.58\textsubscript{.37} & 0.44\textsubscript{.35} & 0.69\textsubscript{.13} & 0.88\textsubscript{.01} & 0.88\textsubscript{.00} & 0.88\textsubscript{.00} & 0.88\textsubscript{.00} \\\midrule
 
 \multirow{6}{*}{\underline{Blood}} & LDP-TabICL& 1 & \textbf{0.76\textsubscript{.07}}& 0.74\textsubscript{.05}& 0.72\textsubscript{.05}& 0.66\textsubscript{.11}& 0.73\textsubscript{.03}& 0.70\textsubscript{.08}& 0.74\textsubscript{.04}\\
                       & LDP-TabICL & 2 & 0.71\textsubscript{.14}& \textbf{0.78\textsubscript{.02}}&0.73\textsubscript{.05}& 0.74\textsubscript{.10}& 0.75\textsubscript{.04}& 0.68\textsubscript{.07}&0.69\textsubscript{.03} \\
                       & LDP-TabICL & 4 & 0.75\textsubscript{.06}& 0.75\textsubscript{.05}& 0.74\textsubscript{.08}& \textbf{0.77\textsubscript{.03}}& 0.57\textsubscript{.22}& 0.67\textsubscript{.12}& \textbf{0.75\textsubscript{.02}}\\
                       & LDP-TabICL & 8 & 0.57\textsubscript{.21}& 0.77\textsubscript{.02}& \textbf{0.78\textsubscript{.04}}& 0.66\textsubscript{.17}& 0.75\textsubscript{.01}& 0.69\textsubscript{.10}& 0.69\textsubscript{.04}\\
                        & LR*        & - & 0.51\textsubscript{.22} & 0.52\textsubscript{.21} & 0.74\textsubscript{.04} & \textbf{0.77\textsubscript{.02}} & \textbf{0.78\textsubscript{.01}} & \textbf{0.77\textsubscript{.04}} & 0.74\textsubscript{.00} \\
                        & GNB*         & - & 0.48\textsubscript{.20} & 0.53\textsubscript{.20} & 0.75\textsubscript{.04} & 0.71\textsubscript{.06} & 0.73\textsubscript{.03} & \textbf{0.77\textsubscript{.02}} & 0.74\textsubscript{.03} \\\midrule
  
 \multirow{6}{*}{\underline{Car}} & LDP-TabICL& 1 & \textbf{0.75\textsubscript{.06}}& \textbf{0.78\textsubscript{.04}}& 0.67\textsubscript{.12}& \textbf{0.72\textsubscript{.04}}& 0.70\textsubscript{.08}& 0.69\textsubscript{.03}& 0.74\textsubscript{.02}\\
                       & LDP-TabICL & 2 & 0.69\textsubscript{.19}& 0.68\textsubscript{.18}& \textbf{0.72\textsubscript{.06}}& 0.71\textsubscript{.08}& 0.72\textsubscript{.04}& 0.77\textsubscript{.09}& 0.75\textsubscript{.02}\\
                       & LDP-TabICL & 4 & 0.67\textsubscript{.08}& 0.63\textsubscript{.19}& 0.70\textsubscript{.08}& \textbf{0.72\textsubscript{.04}}& 0.71\textsubscript{.19}& 0.73\textsubscript{.09}& 0.73\textsubscript{.03}\\
                       & LDP-TabICL & 8 & 0.54\textsubscript{.23}& 0.49\textsubscript{.20}& 0.65\textsubscript{.18}& 0.60\textsubscript{.14}& 0.65\textsubscript{.09}& 0.56\textsubscript{.16}& 0.75\textsubscript{.03}\\
                      & LR*        & - & 0.53\textsubscript{.18} & 0.38\textsubscript{.06} & 0.57\textsubscript{.14} & 0.62\textsubscript{.07} & 0.90\textsubscript{.02} & \textbf{0.94\textsubscript{.01}} & \textbf{0.95\textsubscript{.00}} \\
                      & GNB*         & - & 0.52\textsubscript{.17} & 0.37\textsubscript{.05} & 0.56\textsubscript{.15} & 0.61\textsubscript{.07} & \textbf{0.91\textsubscript{.01}} & 0.88\textsubscript{.01} & 0.85\textsubscript{.01} \\\midrule
 
 \multirow{6}{*}{Calhousing} & LDP-TabICL & 1 & 0.53\textsubscript{.03}& 0.51\textsubscript{.03}& 0.50\textsubscript{.04}& 0.51\textsubscript{.06}& 0.53\textsubscript{.02}& 0.52\textsubscript{.02}& 0.53\textsubscript{.03}\\
                       & LDP-TabICL & 2 & 0.54\textsubscript{.04}& 0.52\textsubscript{.04}& 0.51\textsubscript{.03}& 0.55\textsubscript{.08}& 0.60\textsubscript{.11}& 0.54\textsubscript{.05}& 0.59\textsubscript{.03}\\
                       & LDP-TabICL & 4 & \textbf{0.58\textsubscript{.07}}& \textbf{0.57\textsubscript{.07}}& 0.56\textsubscript{.11}& 0.60\textsubscript{.06}& 0.55\textsubscript{.06}& 0.58\textsubscript{.06}& 0.58\textsubscript{.02}\\
                       & LDP-TabICL & 8 & 0.57\textsubscript{.09}& 0.53\textsubscript{.04}&\textbf{0.62\textsubscript{.11}}& 0.57\textsubscript{.11}& 0.51\textsubscript{.04}& 0.55\textsubscript{.08}& 0.58\textsubscript{.02}\\
                        & LR*        & - & 0.50\textsubscript{.01} & 0.54\textsubscript{.07} & 0.52\textsubscript{.14} & \textbf{0.63\textsubscript{.07}} & \textbf{0.75\textsubscript{.00}} & \textbf{0.75\textsubscript{.00}} & \textbf{0.84\textsubscript{.00}} \\
                        & GNB*         & - & 0.50\textsubscript{.01} & 0.54\textsubscript{.07} & 0.54\textsubscript{.14} & \textbf{0.63\textsubscript{.06}} & 0.74\textsubscript{.01} & 0.73\textsubscript{.01} & 0.73\textsubscript{.01} \\\midrule

 \multirow{6}{*}{Diabetes} & LDP-TabICL& 1 & 0.57\textsubscript{.08}& 0.59\textsubscript{.05}& 0.63\textsubscript{.07}& 0.56\textsubscript{.07}& 0.55\textsubscript{.15}& 0.61\textsubscript{.14}&0.56\textsubscript{.05} \\
                       & LDP-TabICL & 2 & 0.64\textsubscript{.10}& 0.58\textsubscript{.13}& 0.63\textsubscript{.14}& 0.64\textsubscript{.10}& 0.64\textsubscript{.08}& 0.62\textsubscript{.05}& 0.60\textsubscript{.05}\\
                       & LDP-TabICL & 4 & 0.64\textsubscript{.14}& \textbf{0.66\textsubscript{.09}}&\textbf{0.70\textsubscript{.06}} & \textbf{0.65\textsubscript{.06}}& 0.68\textsubscript{.03}& 0.68\textsubscript{.03}& 0.65\textsubscript{.03}\\
                       & LDP-TabICL & 8 & \textbf{0.65\textsubscript{.11}}& 0.62\textsubscript{.05}& 0.61\textsubscript{.10}&0.59\textsubscript{.13} & 0.59\textsubscript{.17}& 0.68\textsubscript{.04}& 0.66\textsubscript{.03}\\
                           & LR*        & - & 0.57\textsubscript{.11} & 0.48\textsubscript{.16} & 0.48\textsubscript{.17} & 0.59\textsubscript{.07} & \textbf{0.72\textsubscript{.02}} & \textbf{0.72\textsubscript{.02}} & \textbf{0.79\textsubscript{.00}} \\
                           & GNB*         & - & 0.58\textsubscript{.12} & 0.50\textsubscript{.15} & 0.47\textsubscript{.16} & 0.60\textsubscript{.06} & 0.70\textsubscript{.02} & 0.71\textsubscript{.02} & 0.71\textsubscript{.03} \\\midrule
 \multirow{6}{*}{Heart} & LDP-TabICL & 1 & 0.57\textsubscript{.09}& 0.57\textsubscript{.14}& 0.61\textsubscript{.06}& \textbf{0.61\textsubscript{.06}}& 0.61\textsubscript{.04}& 0.66\textsubscript{.02}& 0.57\textsubscript{.02} \\
                       & LDP-TabICL & 2 & 0.61\textsubscript{.06}& \textbf{0.59\textsubscript{.08}}& \textbf{0.64\textsubscript{.11}}&0.60\textsubscript{.04} & 0.53\textsubscript{.09}& 0.63\textsubscript{.05}& 0.62\textsubscript{.03}\\
                       & LDP-TabICL & 4 & \textbf{0.62\textsubscript{.04}}& 0.58\textsubscript{.07}& 0.62\textsubscript{.02}& 0.58\textsubscript{.07}& 0.61\textsubscript{.10}& 0.66\textsubscript{.04}& 0.60\textsubscript{.05}\\
                       & LDP-TabICL & 8 & 0.58\textsubscript{.06}& 0.57\textsubscript{.11}& 0.55\textsubscript{.09}& 0.53\textsubscript{.06}& 0.60\textsubscript{.11}& 0.61\textsubscript{.02}& 0.64\textsubscript{.04}\\
                        & LR*         & - & 0.46\textsubscript{.16} & 0.52\textsubscript{.16} & 0.59\textsubscript{.09} & 0.56\textsubscript{.17} & 0.67\textsubscript{.07} & 0.78\textsubscript{.01} & 0.85\textsubscript{.00} \\
                        & GNB*         & - & 0.46\textsubscript{.16} & 0.52\textsubscript{.15} & 0.60\textsubscript{.09} & 0.56\textsubscript{.16} & \textbf{0.71\textsubscript{.05}} & \textbf{0.79\textsubscript{.03}} & \textbf{0.85\textsubscript{.01}} \\\midrule
                        
 \multirow{6}{*}{Jungle} & LDP-TabICL & 1 & 0.50\textsubscript{.03}& 0.50\textsubscript{.02}& 0.51\textsubscript{.02}& 0.49\textsubscript{.02}& 0.51\textsubscript{.02}& 0.53\textsubscript{.03}& 0.50\textsubscript{.01}\\
                       & LDP-TabICL & 2 & 0.53\textsubscript{.03}& 0.52\textsubscript{.06}& 0.51\textsubscript{.03}& 0.54\textsubscript{.04}& 0.51\textsubscript{.04}& 0.48\textsubscript{.02}& 0.50\textsubscript{.02}\\
                       & LDP-TabICL & 4 & \textbf{0.57\textsubscript{.04}}& \textbf{0.53\textsubscript{.06}}& 0.50\textsubscript{.03}& 0.49\textsubscript{.03}& 0.51\textsubscript{.06}& 0.56\textsubscript{.04}& 0.54\textsubscript{.02}\\
                       & LDP-TabICL & 8 & 0.55\textsubscript{.08}& 0.51\textsubscript{.02}& \textbf{0.53\textsubscript{.04}}& 0.56\textsubscript{.04}&0.60\textsubscript{.02} & 0.56\textsubscript{.05}& 0.56\textsubscript{.02}\\
                         & LR*         & - & 0.56\textsubscript{.07} & 0.47\textsubscript{.05} & 0.48\textsubscript{.08} & 0.73\textsubscript{.00} & 0.73\textsubscript{.00} & 0.73\textsubscript{.00} & 0.73\textsubscript{.00} \\
                         & GNB*         & - & 0.56\textsubscript{.07} & 0.47\textsubscript{.04} & 0.47\textsubscript{.08} & \textbf{0.74\textsubscript{.00}} & \textbf{0.74\textsubscript{.00}} & \textbf{0.74\textsubscript{.00}} & \textbf{0.74\textsubscript{.00}} \\\bottomrule
\end{tabular}%
}
\end{table}

\subsubsection{GDP-TabICL}
We report the average accuracy and standard deviation of GDP-TabICL and the two baselines across five runs on all datasets and combinations of $k$ and $\epsilon$ in Table \ref{tab:gdp-tabicl-results-13b}. Similar to the LDP-TabICL case, interesting trends in the results only appear when we analyze the results in relation to if the dataset is balanced or unbalanced. In comparison with LDP-TabICL, on the unbalanced dataset there is a strong trend of accuracy increasing as $\epsilon$ increases, especially for the adult and car datasets. We hypothesize that this is due to GDP-TabICL using differentially private averaging. When the privacy is high ($\epsilon$ is small) the generated differentially private averaged demonstration examples resemble random data that is not representative of the underlying tabular data distribution. However, as $\epsilon$ increases, the noise injected into the averages decreases, meaning that the generated examples more closely align with the underlying data distribution. This trend is not seen as clearly in the balanced datasets since the label distributions are already close to ``random guessing" as there is a fairly even split between the true and false classes. Additionally, similar to the LDP-TabICL case, GDP-TabICL performs best on the unbalanced datasets when only a few number of demonstration examples ($k \geq 2$) are used, while it performs best on the balanced datasets when more demonstration examples ($k \geq 4$) are used and we conjecture it is for the same reason as the trend in LDP-TabICL. We also note that on the balanced datasets the baseline models tend to perform better than any setting of $k$ and $\epsilon$, and that the baseline models tend to perform better on all datasets when $\epsilon=\infty$. However, the results generated by GDP-TabICL are fairly similar (especially in the unbalanced case) and still perform well -- especially considering that our proposed methods only perform ICL over very few demonstration examples and not model training on a full training set. We reiterate our statement from Section \ref{sec:intro} that \textit{the goal of this work is not to produce prompt generation frameworks that outperform traditional classification models in the DP setting, but to evaluate the use of standard DP mechanisms for ensuring privacy guarantees in LLM ICL with tabular data.} In general, these results show that LLMs can utilize GDP techniques like the Laplace mechanism to craft demonstration examples that produce reasonably accurate answers on prompted queries.

\begin{table}[t!]
\centering
\caption{GDP-TabICL average accuracy and standard deviation over 5 runs on Llama-2-13B using differing number of splits $k$ and privacy budget value $\epsilon$. LR: logistic regression, GNB: Gaussian Na\"ive Bayes, \underline{Dataset}: unbalanced dataset.}
\label{tab:gdp-tabicl-results-13b}
\resizebox{.95\columnwidth}{!}{%
\begin{tabular}{@{}ccccHccccc@{}}
\toprule
 & \textbf{} & \multicolumn{7}{c}{$\bm{\epsilon}$} \\
\textbf{Dataset} & \textbf{Method} & $\bm{k}$ & \textbf{1} & \textbf{2} & \textbf{5} & \textbf{10} & \textbf{25} & \textbf{50} & $\bm{\infty}$\\ \midrule
\multirow{6}{*}{\underline{Adult}} & GDP-TabICL & 1 & \textbf{0.72\textsubscript{.05}} & \textbf{0.73\textsubscript{.03}}& \textbf{0.74\textsubscript{.04}}& \textbf{0.76\textsubscript{.02}}& \textbf{0.77\textsubscript{.01}}& \textbf{0.77\textsubscript{.01}}& 0.76\textsubscript{.01}\\
  & GDP-TabICL & 2 & 0.57\textsubscript{.02}& 0.59\textsubscript{.04}& 0.59\textsubscript{.04}& 0.63\textsubscript{.03}& 0.63\textsubscript{.02}& 0.62\textsubscript{.02}& 0.80\textsubscript{.03}\\
  & GDP-TabICL & 4 & 0.61\textsubscript{.05}& 0.63\textsubscript{.02}& 0.66\textsubscript{.03}& 0.65\textsubscript{.03}& 0.67\textsubscript{.03}& 0.67\textsubscript{.02}& 0.80\textsubscript{.01}\\
  & GDP-TabICL & 8 & 0.54\textsubscript{.04}& 0.56\textsubscript{.05}& 0.62\textsubscript{.03}& 0.65\textsubscript{.03}& 0.65\textsubscript{.04}& 0.65\textsubscript{.02}& \textbf{0.83\textsubscript{.01}}\\
 & Diffprivlib LR & - & 0.43\textsubscript{.23}& 0.52\textsubscript{.22}& 0.36\textsubscript{.22}& 0.50\textsubscript{.01}& 0.51\textsubscript{.24}& 0.63\textsubscript{.23}& \textbf{0.83\textsubscript{.01}}\\ 
 & Diffprivlib GNB & - & 0.69\textsubscript{.08}& 0.59\textsubscript{.19}& 0.67\textsubscript{.13}& 0.73\textsubscript{.02}& 0.74\textsubscript{.03}& 0.76\textsubscript{.03}& 0.78\textsubscript{.03}\\ \midrule
 
\multirow{6}{*}{\underline{Bank}}  & GDP-TabICL & 1 & \textbf{0.89\textsubscript{.01}}& \textbf{0.88\textsubscript{.01}}& \textbf{0.88\textsubscript{.01}}& \textbf{0.88\textsubscript{.01}}& \textbf{0.88\textsubscript{.01}}& \textbf{0.88\textsubscript{.01}}& 0.88\textsubscript{.01}\\
  & GDP-TabICL & 2 & 0.84\textsubscript{.06}& 0.87\textsubscript{.04}& \textbf{0.88\textsubscript{.01}}& 0.87\textsubscript{.02}& 0.86\textsubscript{.01}& 0.86\textsubscript{.01}& \textbf{0.89\textsubscript{.01}}\\
  & GDP-TabICL & 4 & 0.81\textsubscript{.04}& 0.81\textsubscript{.02}& 0.79\textsubscript{.05}& 0.76\textsubscript{.04}& 0.78\textsubscript{.01}&0.77\textsubscript{.02} & 0.88\textsubscript{.01}\\
  & GDP-TabICL & 8 & 0.66\textsubscript{.13}& 0.62\textsubscript{.08}& 0.64\textsubscript{.14}& 0.64\textsubscript{.06}& 0.61\textsubscript{.09}& 0.61\textsubscript{.12}& 0.88\textsubscript{.00}\\
& Diffprivlib LR & - & 0.47\textsubscript{.26}& 0.41\textsubscript{.25}& 0.49\textsubscript{.03}& 0.52\textsubscript{.26}& 0.39\textsubscript{.21}& 0.50\textsubscript{.38}& \textbf{0.89\textsubscript{.01}}\\ 
 & Diffprivlib GNB & - & 0.84\textsubscript{.07}& 0.78\textsubscript{.12}& 0.83\textsubscript{.04}& 0.85\textsubscript{.02}& 0.84\textsubscript{.03}& 0.83\textsubscript{.02}& 0.82\textsubscript{.03} \\ \midrule
 
 \multirow{6}{*}{\underline{Blood}} & GDP-TabICL & 1 & \textbf{0.80\textsubscript{.03}}& \textbf{0.81\textsubscript{.02}}& \textbf{0.82\textsubscript{.01}}& \textbf{0.82\textsubscript{.02}}& \textbf{0.79\textsubscript{.05}}& \textbf{0.80\textsubscript{.05}}& 0.64\textsubscript{.05}\\
  & GDP-TabICL & 2 & 0.69\textsubscript{.12}& 0.69\textsubscript{.10}& 0.70\textsubscript{.02}& 0.71\textsubscript{.05}& 0.70\textsubscript{.03}& 0.69\textsubscript{.03}& 0.78\textsubscript{.02}\\
  & GDP-TabICL & 4 & 0.75\textsubscript{.11}& 0.64\textsubscript{.09}& 0.67\textsubscript{.06}& 0.69\textsubscript{.04}& 0.70\textsubscript{.02}& 0.69\textsubscript{.04}& 0.75\textsubscript{.04}\\
  & GDP-TabICL & 8 & 0.71\textsubscript{.10}& 0.71\textsubscript{.11}& 0.65\textsubscript{.01}& 0.66\textsubscript{.07}& 0.64\textsubscript{.09}& 0.65\textsubscript{.10}& 0.68\textsubscript{.10}\\
 & Diffprivlib LR & - & 0.43\textsubscript{.28}& 0.45\textsubscript{.27}& 0.43\textsubscript{.27}& 0.57\textsubscript{.28}& 0.52\textsubscript{.28}& 0.51\textsubscript{.30}& \textbf{0.83\textsubscript{.03}}\\ 
 & Diffprivlib GNB & - & 0.77\textsubscript{.08}& 0.75\textsubscript{.11}& 0.78\textsubscript{.03}& 0.78\textsubscript{.02}& 0.76\textsubscript{.08}& \textbf{0.80\textsubscript{.02}}& 0.82\textsubscript{.03}\\ \midrule
  
 \multirow{6}{*}{\underline{Car}}  & GDP-TabICL & 1 & \textbf{0.73\textsubscript{.04}}& \textbf{0.78\textsubscript{.03}}& 0.75\textsubscript{.05}& 0.78\textsubscript{.02}& 0.79\textsubscript{.02}& 0.78\textsubscript{.02}& 0.75\textsubscript{.06}\\
  & GDP-TabICL & 2 & 0.68\textsubscript{.08}& 0.66\textsubscript{.02}& 0.69\textsubscript{.02}& 0.69\textsubscript{.02}& 0.70\textsubscript{.05}& 0.69\textsubscript{.01}& 0.69\textsubscript{.01}\\
  & GDP-TabICL & 4 & 0.63\textsubscript{.08}& 0.66\textsubscript{.04}& 0.68\textsubscript{.02}& 0.67\textsubscript{.02}& 0.69\textsubscript{.02}& 0.66\textsubscript{.03}& 0.70\textsubscript{.04}\\
  & GDP-TabICL & 8 & 0.57\textsubscript{.07}& 0.65\textsubscript{.11}& 0.65\textsubscript{.06}& 0.71\textsubscript{.04}& 0.69\textsubscript{.05}& 0.67\textsubscript{.06}& 0.70\textsubscript{.06}\\
& Diffprivlib LR & - & 0.46\textsubscript{.15}& 0.52\textsubscript{.13}& 0.56\textsubscript{.10}& 0.52\textsubscript{.10}& 0.58\textsubscript{.18}&0.75\textsubscript{.06} & \textbf{0.94\textsubscript{.01}}\\ 
 & Diffprivlib GNB & - & 0.66\textsubscript{.09}& 0.72\textsubscript{.13}& \textbf{0.84\textsubscript{.08}}& \textbf{0.82\textsubscript{.09}}& \textbf{0.89\textsubscript{.02}}& \textbf{0.88\textsubscript{.02}}& 0.85\textsubscript{.01}\\ \midrule
 
 \multirow{6}{*}{Calhousing}  & GDP-TabICL & 1 & 0.59\textsubscript{.04}& 0.56\textsubscript{.06} & 0.57\textsubscript{.05}& 0.59\textsubscript{.06}& 0.57\textsubscript{.04}& 0.61\textsubscript{.05}& 0.54\textsubscript{.06}\\
  & GDP-TabICL & 2 & 0.56\textsubscript{.03}& 0.53\textsubscript{.02}& 0.55\textsubscript{.03}& 0.55\textsubscript{.03}& 0.52\textsubscript{.01}& 0.53\textsubscript{.02}& 0.48\textsubscript{.01}\\
  & GDP-TabICL & 4 & 0.61\textsubscript{.07}& 0.61\textsubscript{.07}& 0.60\textsubscript{.06}& 0.61\textsubscript{.08}& 0.57\textsubscript{.04}& 0.58\textsubscript{.07}& 0.63\textsubscript{.03}\\
  & GDP-TabICL & 8 & \textbf{0.65\textsubscript{.06}}& \textbf{0.64\textsubscript{.05}}& 0.61\textsubscript{.05}& 0.64\textsubscript{.04}& 0.64\textsubscript{.03}& 0.64\textsubscript{.07}& 0.61\textsubscript{.03}\\
& Diffprivlib LR & - & 0.50\textsubscript{.06}& 0.53\textsubscript{.06}& 0.47\textsubscript{.04}& 0.50\textsubscript{.05}& 0.51\textsubscript{.05}& 0.52\textsubscript{.02}& \textbf{0.77\textsubscript{.03}}\\ 
 & Diffprivlib GNB & - & 0.57\textsubscript{.11}& 0.58\textsubscript{.08}& \textbf{0.69\textsubscript{.07}}& \textbf{0.69\textsubscript{.07}}& \textbf{0.69\textsubscript{.06}}& \textbf{0.70\textsubscript{.06}}& 0.72\textsubscript{.02}\\ \midrule

 \multirow{6}{*}{Diabetes}  & GDP-TabICL & 1 & 0.64\textsubscript{.04}& 0.59\textsubscript{.04}& 0.63\textsubscript{.03}& 0.57\textsubscript{.05}& 0.57\textsubscript{.05}& 0.54\textsubscript{.03}& 0.55\textsubscript{.02}\\
  & GDP-TabICL & 2 &\textbf{0.71\textsubscript{.08}}& \textbf{0.75\textsubscript{.04}}& 0.74\textsubscript{.03}& \textbf{0.76\textsubscript{.04}}& 0.77\textsubscript{.04}&\textbf{0.78\textsubscript{.02}} & 0.71\textsubscript{.04}\\
  & GDP-TabICL & 4 & 0.68\textsubscript{.04}& 0.74\textsubscript{.03}& \textbf{0.75\textsubscript{.06}}& 0.70\textsubscript{.05}& 0.73\textsubscript{.04}& 0.73\textsubscript{.02}& 0.73\textsubscript{.05}\\
  & GDP-TabICL & 8 & 0.63\textsubscript{.07}& 0.60\textsubscript{.06}& 0.63\textsubscript{.06}& 0.62\textsubscript{.09}& 0.62\textsubscript{.06}& 0.66\textsubscript{.05}& 0.69\textsubscript{.02}\\
& Diffprivlib LR & - & 0.42\textsubscript{.13}& 0.51\textsubscript{.16}& 0.54\textsubscript{.14}& 0.42\textsubscript{.11}& 0.47\textsubscript{.15}& 0.45\textsubscript{.17}& 0.77\textsubscript{.01}\\ 
 & Diffprivlib GNB & - & 0.61\textsubscript{.08}& 0.69\textsubscript{.04}& 0.71\textsubscript{.05}& 0.68\textsubscript{.08}& \textbf{0.78\textsubscript{.03}}& 0.77\textsubscript{.03}& \textbf{0.79\textsubscript{.02}}\\ \midrule
 
 \multirow{6}{*}{Heart}  & GDP-TabICL & 1 & 0.64\textsubscript{.05}& 0.64\textsubscript{.04}& 0.64\textsubscript{.03}& 0.64\textsubscript{.02}& 0.64\textsubscript{.03}& 0.67\textsubscript{.01}& 0.67\textsubscript{.02}\\
  & GDP-TabICL & 2 & 0.72\textsubscript{.04}& 0.72\textsubscript{.03}& 0.70\textsubscript{.02}& 0.70\textsubscript{.07}& 0.69\textsubscript{.04}& 0.72\textsubscript{.04}& 0.71\textsubscript{.03}\\
  & GDP-TabICL & 4 & \textbf{0.76\textsubscript{.03}}& \textbf{0.78\textsubscript{.03}}& \textbf{0.77\textsubscript{.02}}& 0.74\textsubscript{.02}& 0.76\textsubscript{.02}& 0.74\textsubscript{.02}& 0.75\textsubscript{.03}\\
  & GDP-TabICL & 8 & 0.72\textsubscript{.04}& 0.73\textsubscript{.04}& 0.72\textsubscript{.03}& 0.74\textsubscript{.04}& 0.73\textsubscript{.02}& 0.72\textsubscript{.02}& 0.69\textsubscript{.03}\\
& Diffprivlib LR & - & 0.47\textsubscript{.06}&0.49\textsubscript{.07}& 0.50\textsubscript{.08}& 0.48\textsubscript{.07}& 0.54\textsubscript{.08}& 0.54\textsubscript{.05}& \textbf{0.86\textsubscript{.03}}\\ 
 & Diffprivlib GNB & - & 0.74\textsubscript{.08}& 0.74\textsubscript{.11}& 0.76\textsubscript{.10}& \textbf{0.78\textsubscript{.09}}& \textbf{0.84\textsubscript{.02}}& \textbf{0.85\textsubscript{.02}}& 0.84\textsubscript{.02}\\ \midrule
 
 \multirow{6}{*}{Jungle}  & GDP-TabICL & 1 & 0.53\textsubscript{.02}& 0.52\textsubscript{.02}& 0.51\textsubscript{.01}& 0.53\textsubscript{.01}& 0.53\textsubscript{.01}& 0.52\textsubscript{.00}& 0.48\textsubscript{.01}\\
  & GDP-TabICL & 2 &0.51\textsubscript{.01} & 0.51\textsubscript{.01}& 0.52\textsubscript{.01}& 0.52\textsubscript{.01}& 0.52\textsubscript{.02}& 0.52\textsubscript{.01}& 0.48\textsubscript{.01}\\
  & GDP-TabICL & 4 & 0.51\textsubscript{.01}& 0.51\textsubscript{.01}& 0.52\textsubscript{.02}& 0.52\textsubscript{.01}& 0.52\textsubscript{.00}& 0.51\textsubscript{.01}& 0.50\textsubscript{.02}\\
  & GDP-TabICL & 8 & 0.59\textsubscript{.04}& 0.59\textsubscript{.03}& 0.58\textsubscript{.04}& 0.61\textsubscript{.02}& 0.60\textsubscript{.01}& 0.60\textsubscript{.01}& 0.61\textsubscript{.01}\\
& Diffprivlib LR & - & 0.50\textsubscript{.05}& 0.51\textsubscript{.04}& 0.45\textsubscript{.05}& 0.52\textsubscript{.06}& 0.52\textsubscript{.02}& 0.73\textsubscript{.02}& 0.73\textsubscript{.01}\\ 
 & Diffprivlib GNB & - & \textbf{0.70\textsubscript{.07}}& \textbf{0.71\textsubscript{.08}}& \textbf{0.75\textsubscript{.01}}& \textbf{0.75\textsubscript{.02}}& \textbf{0.75\textsubscript{.01}}& \textbf{0.75\textsubscript{.01}}& \textbf{0.75\textsubscript{.01}}\\ 
 \bottomrule
\end{tabular}%
}
\end{table}

\begin{figure}[t!]
     \centering
    \begin{subfigure}[t]{0.49\columnwidth}
        \centering
        \includegraphics[width=\columnwidth]{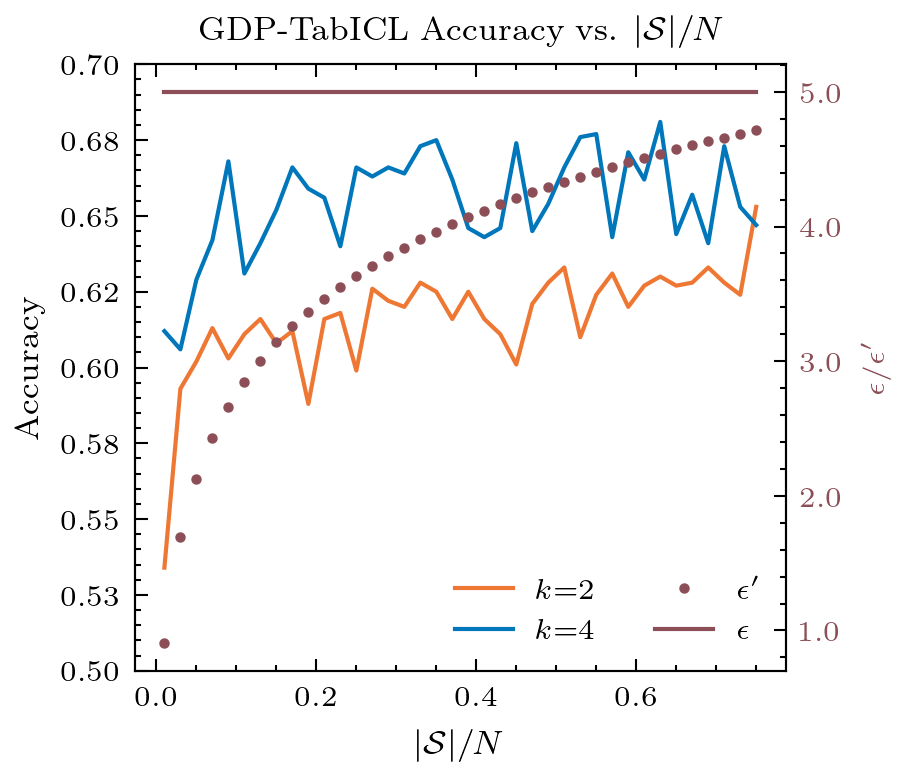}
        \caption{Adult}
    \end{subfigure}
    \begin{subfigure}[t]{0.49\columnwidth}
        \centering
        \includegraphics[width=\columnwidth]{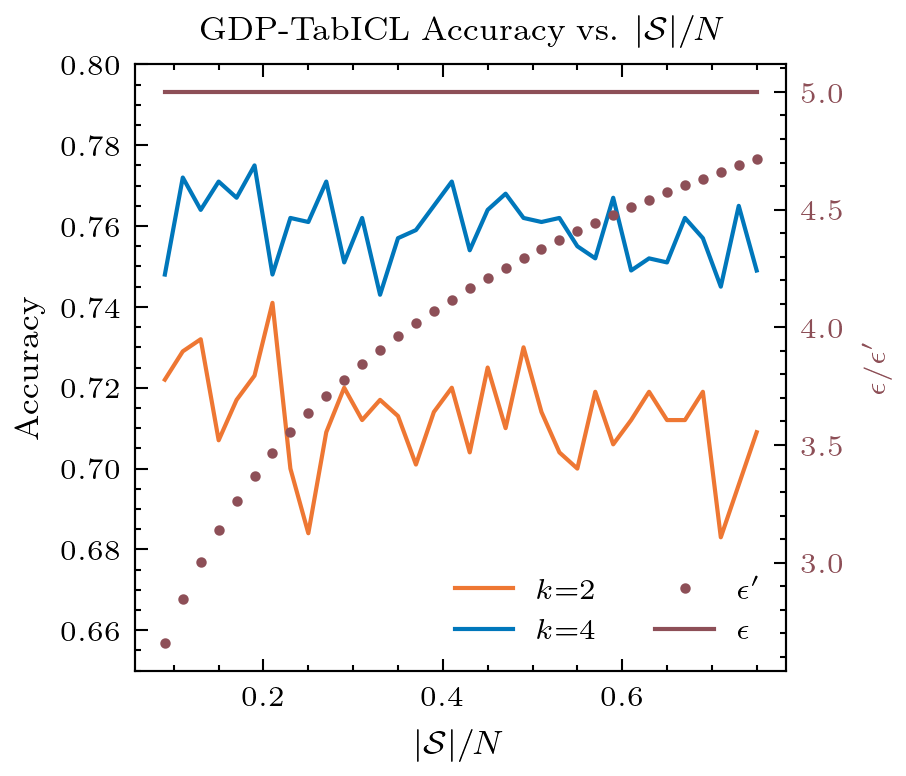}
        \caption{Heart}
    \end{subfigure}
    \caption{Accuracy of GDP-TabICL when the size of $\mathcal{S}$ is altered for the Adult (top) and Heart (bottom) datasets with $\epsilon=5$.}
    \label{fig:gdp-tabicl-ablation}
\end{figure}

\paragraph*{\textbf{Ablation on Size of Subset} $\bm{\mathcal{S}}$}
Here, we study how the size $n = |\mathcal{S}|$ of the subset $\mathcal{S}$, selected using Poisson sampling affects the overall accuracy of an LLM on differentially private tabular data in GDP-TabICL. For both the adult and heart dataset, we select $k \in \{2, 4\}$ and $\epsilon = 5$ and let $n = |\mathcal{S}|$ range from 1\%-75\% the size of the total dataset size $N$. We show results for all cases in Fig. \ref{fig:gdp-tabicl-ablation}. On the adult dataset, there is a clear increase in performance when the size of $\mathcal{S}$ increase above $2\%$. This is because the subset size is too small to contain a good representation of the overall dataset due to the unbalanced nature of the adult data. However, when the size of $\mathcal{S}$ allows an adequate representation of the underlying dataset, then increasing the size of $\mathcal{S}$ only has a slight effect on the overall accuracy (or in the case of the heart dataset, no observable trend). This is because we only send the differentially private aggregates as demonstration examples, and since the number of demonstration examples does not change, neither does the accuracy. 

\subsubsection{LDP-TabICL vs. GDP-TabICL}
While in this work we simply propose two differentially private methods to protect tabular data used for in-context learning, and do not necessarily aim to say that one proposed method is better than the other, there are noticeable differences between when the two methods perform best. Specifically, LDP-TabICL works best on unbalanced datasets -- except in the case when only one demonstration example is used ($k=1$), or when $\epsilon=\infty$. We conjecture that this is due to the demonstration examples from GDP-TabICL containing richer information of the underlying dataset via using differentially private averages while LDP-TabICL randomly samples a single LDP-protected (and reconstructed) demonstration example that may or may not be representative of the dataset. In contrast, GDP-TabICL performs best on balanced datasets. While we attempted to make demonstration examples that represented the underlying dataset by using the GROUP BY query to create the partition of $\mathcal{S}$, using averages naturally favors using balanced data as using unbalanced data can cause classes that are not well represented to not appear in the DP averaged demonstration examples. Therefore, we recommend using LDP-TabICL for unbalanced datasets with $k\geq2$ and using GDP-TabICL on balanced datasets (or on unbalanced datasets when $k=1$).

\subsubsection{Llama-2-13B vs. Llama-2-7B}
In this work, we ran all experiments on the Llama-2-7B and Llama-2-13B models. To see if the chosen model size had a significant impact on the results, we performed a two-tailed paired Student t-test between the 7B and 13B results. We report the $p$-values for each $k$ value (taken over all $\epsilon$ values) for the LDP-TabICL and GDP-TabICL and in Table \ref{tab:gdp-pvalue}. For LDP-TabICL, half of the $p$-values are well above 0.05 implying that the model choice generally does not significantly impact LLM inference when data is perturbed locally. However, the results from the t-test with GDP-TabICL do not follow the same trend. In all but 2 settings (blood $k=4$ and diabetes $k=8$) the resulting $p$-value was well below 0.05 meaning that the model size had a statistically significant impact on test performance when GDP mechanisms were implemented.

\begin{table}[t!]
\centering
\caption{$p$-value from paired student t-test on LDP-TabICL (LDP) and GDP-TabICL (GDP) results from Llama-2-7B and Llama-2-13B. Bold values $\geq 0.05$.}
\label{tab:gdp-pvalue}
\resizebox{\columnwidth}{!}{%
\begin{tabular}{@{}cccccccccc@{}}
\toprule
 & \multicolumn{9}{c}{\textbf{Dataset}} \\ 
\textbf{Method} & $\bm{k}$ & \textbf{Adult} & \textbf{Bank} & \textbf{Blood} & \textbf{Calhousing} & \textbf{Car} & \textbf{Diabetes} & \textbf{Heart} & \textbf{Jungle} \\\midrule
\multirow{4}{*}{LDP} & 1 &  \textbf{0.447} & 0.007 & \textbf{0.170} & 0.005 & 0.003 & 0.002& \textbf{0.086} & \textbf{0.876}\\
& 2 & \textbf{0.539} & \textbf{0.799} & 0.004 & 0.027& \textbf{0.917} & 0.010& 0.002 & 0.037\\
& 4 & \textbf{0.134} & \textbf{0.125} & \textbf{0.185} & 0.000 & \textbf{0.060} & 0.033&  0.001& 0.024\\
& 8 & 0.025 & \textbf{0.347} & \textbf{0.407} & \textbf{0.124} & 0.016 & \textbf{0.219}&\textbf{0.286} & 0.004\\\midrule
\multirow{4}{*}{GDP} & 1 & 0.051 & 0.030 & 0.001 & 0.014 & 0.007 & 0.004 & 0.000 & 0.003 \\
& 2 & 0.002 & 0.013 & 0.001 & 0.009 & 0.000 & 0.003 & 0.000 & 0.003 \\
& 4 & 0.002 & 0.002 & \textbf{0.482} & 0.007 & 0.000 & 0.030 & 0.002 & 0.001 \\
& 8 & 0.002 & 0.001 & 0.002 & 0.005 & 0.009 & \textbf{0.341} & 0.000 & 0.000 \\ \bottomrule
\end{tabular}%
}
\end{table}

\section{Conclusion}
\label{sec:conclusion}
In this work, we presented and evaluated LDP-TabICL and GDP-TabICL -- two frameworks to produce differentially private demonstrations for in-context learning on tabular data. For both LDP-TabICL and GDP-TabICL, we utilized standard differentially private mechanisms and, specifically, we used randomized response in the formulation of LDP-TabICL and the Laplace mechanism in the formulation of GDP-TabICL. The main idea behind LDP-TabICL is to introduce randomization for individual records, while GDP-TabICL introduces additive noise into the feature/label averages. Further, we also incorporate frequency estimation as post-processing (LDP-TabICL) and sampling strategies as pre-processing (GDP-TabICL) to obtain perturbed data that approximately represents the feature and label distribution of the underlying private tabular data. We ultimately used the perturb data to construct LLM prompts via serialization with a manual text template. Experimental evaluation on eight real-world tabular datasets, across multiple ICL and DP settings, demonstrated the viability of using standard LDP and GDP techniques to protect tabular data used for ICL, especially under strict privacy requirements. Future work will entail the exploration of prompt optimization and instruction-tuning strategies to achieve better utility-privacy trade-offs as well as the implementation of different LDP and GDP mechanisms for improved estimation.   

\section*{Ethics Statement}
Using LLMs inherently brings into conversation the topic of ethics, especially in relation to the data the model is trained on. We acknowledge that the data used to train the models evaluated in this work are inherently non-private and that simply protecting the underlying tabular data used in ICL only does so much in increasing the privacy of the overall LLM pipeline. Further, we note that using privacy protecting mechanisms on the demonstration examples can potentially produce unwanted effects on the fairness of the generated LLM results, and we plan to perform future work along this direction. Additionally, we acknowledge the environmental implications of the use of LLMs and provide details regarding the CO\textsubscript{2} consumption of conducted experiments in Appendix \ref{app:comp-cost}.

\section*{Reproducibility}
In the effort to keep research open and reproducible, we publish all of our source code at \url{http://tinyurl.com/remyzjxk}. 

\section*{Acknowledgements}
This work was supported in part by the National Science Foundation under awards 1920920, 1946391, and 2119691, the National Institute of General Medical Sciences of National Institutes of Health under award P20GM139768, and the Arkansas Integrative Metabolic Research Center at the University of Arkansas.

\bibliographystyle{ieeetr}
\bibliography{references}
\clearpage
\begin{appendix}

\subsection{Additional Experimental Settings}
\subsubsection{Privacy Amplification via Subsampling}
In Table \ref{tab:e-vs-e'} we report the resulting change in the privacy parameter $\epsilon$ when performing subsampling in GDP-TabICL. 
\label{app:amp-by-sub}
\begin{table}[h!]
\centering
\caption{$\epsilon$ vs. $\epsilon'=\ln(1+\frac{n}{N}(e^\epsilon - 1))$ for all datasets.}
\label{tab:e-vs-e'}
\resizebox{.9\columnwidth}{!}{%
\begin{tabular}{@{}ccccccccc@{}}
\toprule
 & \multicolumn{8}{c}{$\bm{\epsilon'}$} \\
\textbf{} & \textbf{Adult} & \textbf{Bank} & \textbf{Blood} & \textbf{Calhousing} & \textbf{Car} & \textbf{Diabetes} & \textbf{Heart} & \textbf{Jungle} \\
$\bm{\epsilon}$ & \begin{tabular}[c]{@{}c@{}}n = 3016\\ N = 30162\end{tabular} & \begin{tabular}[c]{@{}c@{}}n = 3617\\ N = 36168\end{tabular} & \begin{tabular}[c]{@{}c@{}}n = 300\\ N = 598\end{tabular} & \begin{tabular}[c]{@{}c@{}}n = 3300\\ N = 16512\end{tabular} & \begin{tabular}[c]{@{}c@{}}n = 691\\ N = 1382\end{tabular} & \begin{tabular}[c]{@{}c@{}}n = 307\\ N = 617\end{tabular} & \begin{tabular}[c]{@{}c@{}}n = 367\\ N = 734\end{tabular} & \begin{tabular}[c]{@{}c@{}}n = 3586\\ N = 35855\end{tabular} \\ \midrule
1 & 0.159 & 0.159 & 0.622 & 0.295 & 0.620 & 0.620 & 0.620 & 0.159 \\
5 & 2.756 & 2.756 & 4.317 & 4.416 & 4.314 & 4.314 & 4.314 & 2.756 \\
10 & 7.698 & 7.698 & 9.310 & 8.390 & 9.307 & 9.307 & 9.307 & 7.698 \\
25 & 22.697 & 22.697 & 24.310 & 23.390 & 24.307 & 24.307 & 24.307 & 22.698 \\
50 & 47.697 & 47.370 & 49.310 & 48.390 & 49.307 & 49.307 & 49.307 & 47.698
\end{tabular}
}
\end{table}

\subsubsection{Dataset Descriptions}
In Table \ref{tab:train-test-size} we report the size of the training set $N_{tr}$ and testing set $N_{te}$ for each dataset. The original dataset was split 80/20 to form the train and test sets. However, due to computational constraints, the entire test set was not used during experimentation. Instead, either a 10\% (adult, bank, calhousing, jungle) or 75\% (blood, car, diabetes, heart) subset of the test set was used. 
\begin{table}[h!]
\centering
\caption{Size of train $N_{tr}$ and test $N_{te}$ sets for all datasets.}
\label{tab:train-test-size}
\resizebox{.9\columnwidth}{!}{%
\begin{tabular}{@{}ccccccccc@{}}
\toprule
 & \multicolumn{8}{c}{\textbf{Dataset}} \\ 
 & \textbf{Adult} & \textbf{Bank} & \textbf{Blood} & \textbf{Calhousing} & \textbf{Car} & \textbf{Diabetes} & \textbf{Heart} & \textbf{Jungle} \\ \midrule
$N_{tr}$ & 30162 & 36168 & 598 & 16512 & 1382 & 614 & 734 & 35855 \\
$N_{te}$ & 1506 & 904 & 112 & 412 & 259 & 115 & 138 & 896 \\ \bottomrule
\end{tabular}%
}
\end{table}

\subsubsection{GROUP BY}
\label{app:group-by}
In this work, we use the GROUP BY method to split $\mathcal{S}$ into $k$ subsets to form a partition of $\mathcal{S}$. For each dataset, we craft 4 different GROUP BY queries listed in Table \ref{tab:group-by}. 
When $k=1$, we do not perform GROUP BY and simply take the DP average of $\mathcal{S}$ in its entirety. Further, we assume that each split is non-empty.

\begin{table}[h!]
\centering
\caption{GROUP BY columns:values for generating splits.}
\label{tab:group-by}
\resizebox{.9\columnwidth}{!}{%
\begin{tabular}{@{}c|l|l@{}}
\toprule
Dataset & \multicolumn{1}{c|}{$k$} & \multicolumn{1}{c|}{GROUP BY Column(s) : Values} \\ \midrule
\multirow{3}{*}{Adult} 
 & 2 & Label: [True, False] \\
 & 4 & Label, Sex: [True, False], [Male, Female] \\
 & 8 & Label, Sex, Race: [True, False], [Male, Female], [White, Non-White]\\\midrule
 
\multirow{3}{*}{Bank} 
 & 2 & Label: [True, False] \\
 & 4 & Label, Marital: [True, False], [Married, Not-Married]\\
 & 8 & Label, Marital: [True, False], [Married, Not-Married], [Cellular, Non-Cellular]\\\midrule
 
 \multirow{3}{*}{Blood} 
 & 2 & Labe: [True, False] \\
 & 4 & Label, Frequency: [True, False], [$\leq 5$, $>5$] \\
 & 8 & Label, Frequency, Recency: [True, False], [$\leq 5$, $>5$], [$\leq10$, $>10$]\\ \midrule

 \multirow{3}{*}{Calhousing} 
 & 2 & Label: [True, False] \\
 & 4 & Label, Housing Median Age: [True, False], [$\leq25$, $>25$]  \\
 & 8 & Label, Housing Median Age, Population: [True, False], [$\leq25$, $>25$], [$\leq2000$, $>2000$]\\ \midrule

 \multirow{3}{*}{Car} 
 & 2 & Label: [True (1, 2, 3), False (0)]\\
 & 4 & Label, Buying: [True (1, 2, 3), False (0)], [(low, med), (high, vhigh)]\\
 & 8 & Label, Buying, Doors: [True (1, 2, 3), False (0)], [(low, med), (high, vhigh)], [$<4$, $\geq4$]\\ \midrule

 \multirow{3}{*}{Diabetes} 
 & 2 & Label: [True, False] \\
 & 4 & Label, Pregnancies: [True, False], [$\leq4$, $>4$] \\
 & 8 & Label, Pregnancies, Age: [True, False], [$\leq4$, $>4$], [$\leq33$, $>33$]\\ \midrule

 \multirow{3}{*}{Heart} 
 & 2 & Label: [True, False] \\
 & 4 & Label, Sex: [True, False], [M, F] \\
 & 8 & Label, Sex, ExerciseAngina: [True, False], [M, F], [N, Y]\\ \midrule

 \multirow{3}{*}{Jungle} 
 & 2 & Label: [True, False] \\
 & 4 & Label, White Strength: [True, False], [$\leq4$, $>4$] \\
 & 8 & Label, White Strength, Black Strength: [True, False], [$\leq4$, $>4$], [$\leq4$, $>4$]\\ 
 \bottomrule
\end{tabular}%
}
\end{table}

\subsubsection{Example Prompts}
\label{app:examples}
We note that no prompt engineering was completed in this work, and more optimal prompts for both the LDP-TabICL and GDP-TabICL cases may exist. We further note that while we set $k=2$ and $\epsilon=5$ to generate the prompts, the format is the same regardless of the chosen $k$ or $\epsilon$ value.
\paragraph{\textbf{LDP}} In Table \ref{tab:prompt-examples-ldp} we show prompt examples generated for the LDP-TabICL setting for each dataset. The prompt templates were crafted specifically for each dataset, and since the datasets for the LDP case were made to be binary, the statements in the description and query consist of phrases such as ``more than", ``less than or equal to", and ``at most" rather than concrete phrases such as ``39 year old male." 
\begin{table*}[h!]
\centering
\caption{Example prompts for LDP-TabICL when $k=2$ and $\epsilon=5$.}
\label{tab:prompt-examples-ldp}
\resizebox{.9\textwidth}{!}{%
\begin{tabular}{@{}c|cl@{}}
\toprule
Dataset & \multicolumn{2}{c|}{Prompt} \\ \midrule
\multirow{2}{*}{Adult} & \multicolumn{1}{c|}{Demonstration} & \begin{tabular}[c]{@{}l@{}}An individual recorded in the 1994 US census is described as follows: This person is a male 39 or more years of age. He has a high school degree at most.\\ He does not work in the private sector. He works less than 40 hours per week. His capital gain was more than \$1092 last year. He is White. He is not married. \\ Does this person earn more than \$50,000 dollars annually? Yes or No? Answer: No\\ \\ An individual recorded in the 1994 US census is described as follows: This person is a male less than 39 years of age. He has an associate's degree at most.\\ He works in the private sector. He works 40 or more hours per week. His capital gain was more than \$1092 last year. He is White. He is married. Does this \\ person earn more than \$50,000 dollars annually? Yes or No? Answer: Yes\end{tabular} \\ \cmidrule(l){2-3} 
 & \multicolumn{1}{c|}{Query} & \begin{tabular}[c]{@{}l@{}}An individual recorded in the 1994 US census is described as follows: This person is a female 39 or more years of age. She has a college degree at most.\\ She works in the private sector. She works less than 40 hours per week. Her capital gain was less than \$1092 last year. She is White. She is married. Does\\ this person earn more than \$50,000 dollars annually? Yes or No? Answer:\end{tabular} \\ \midrule
\multirow{2}{*}{Bank} & \multicolumn{1}{c|}{Demonstration} & \begin{tabular}[c]{@{}l@{}}A client at a Portuguese banking institution is described as follows: The client is 40 or less years of age. Their average yearly balance is less than 1362 euros.\\ They have housing loans. They were contacted less than 3 times during this campaign and were last contacted for less than 258 seconds. They were contacted\\ one or more times in a previous campaign and were last contacted more than 40 days ago for the previous campaign. The outcome of the previous marketing\\ campaign was either failure or unknown for this client. Does this person subscribe to a term deposit? Yes or No? Answer: Yes\\ \\ A client at a Portuguese banking institution is described as follows: The client is more than 40 years of age. Their average yearly balance is more than 1362 \\ euros. They have housing loans. They were contacted less than 3 times during this campaign and were last contacted for more than 258 seconds. They were \\ contacted one or more times in a previous campaign and were last contacted more than 40 days ago for the previous campaign. The outcome of the previous\\ marketing campaign was either failure or unknown for this client. Does this person subscribe to a term deposit? Yes or No? Answer: Yes\end{tabular} \\ \cmidrule(l){2-3} 
 & \multicolumn{1}{c|}{Query} & \begin{tabular}[c]{@{}l@{}}A client at a Portuguese banking institution is described as follows: The client is 40 or less years of age. Their average yearly balance is less than 1362 euros.\\ They have housing loans. They were contacted less than 3 times during this campaign and were last contacted for less than 258 seconds. They were not \\ contacted in previous campaigns. The outcome of the previous marketing campaign was either failure or unknown for this client. Does this person subscribe\\  to a term deposit? Yes or No? Answer:\end{tabular} \\ \midrule
\multirow{2}{*}{Blood} & \multicolumn{1}{c|}{Demonstration} & \begin{tabular}[c]{@{}l@{}}A blood donor at a Blood Transfusion Service Center is described as follows: The donor has donated blood 6 or more times. In total, they have donated less \\ than 1379 c.c. of blood. They last donated blood 10 or more months ago. Their first blood donation was more than 34 months ago. Did the donor donate \\ blood in March 2007? Yes or No? Answer: Yes\\ \\ A blood donor at a Blood Transfusion Service Center is described as follows: The donor has donated blood 6 or more times. In total, they have donated less\\ than 1379 c.c. of blood. They last donated blood 10 or more months ago. Their first blood donation was less than 34 months ago. Did the donor donate \\ blood in March 2007? Yes or No? Answer: No\end{tabular} \\ \cmidrule(l){2-3} 
 & \multicolumn{1}{c|}{Query} & \begin{tabular}[c]{@{}l@{}}A blood donor at a Blood Transfusion Service Center is described as follows: The donor has donated blood 6 or more times. In total, they have donated more \\ than 1379 c.c. of blood. They last donated blood less than 10 months ago. Their first blood donation was more than 34 months ago. Did the donor donate \\ blood in March 2007? Yes or No? Answer:\end{tabular} \\ \midrule
\multirow{2}{*}{Calhousing} & \multicolumn{1}{c|}{Demonstration} & \begin{tabular}[c]{@{}l@{}}A house block in California has the following attributes according to the 1990 California census. This housing block is located at latitude more than 36 and \\ longitude less than -120. The houses in the block have more than 2636 rooms with more than 538 bedrooms in total. The median age of houses in the block \\ is less than 29 years. There are more than 500 total households in the block with a total population more than 1425. The median income of the households \\ in the block is more than 40,000 dollars. Is this housing block valuable? Yes or No? Answer: No\\ \\ A house block in California has the following attributes according to the 1990 California census. This housing block is located at latitude more than 36 and\\ longitude more than -120. The houses in the block have more than 2636 rooms with less than 538 bedrooms in total. The median age of houses in the block\\ is less than 29 years. There are less than 500 total households in the block with a total population more than 1425. The median income of the households \\ in the block is less than 40,000 dollars. Is this housing block valuable? Yes or No? Answer: Yes\end{tabular} \\ \cmidrule(l){2-3} 
 & \multicolumn{1}{c|}{Query} & \begin{tabular}[c]{@{}l@{}}A house block in California has the following attributes according to the 1990 California census. This housing block is located at latitude more than 36 and \\ longitude less than -120. The houses in the block have less than 2636 rooms with less than 538 bedrooms in total. The median age of houses in the block is \\ more than 29 years. There are less than 500 total households in the block with a total population less than 1425. The median income of the households in the \\ block is less than 40,000 dollars. Is this housing block valuable? Yes or No? Answer:\end{tabular} \\ \midrule
\multirow{2}{*}{Car} & \multicolumn{1}{c|}{Demonstration} & \begin{tabular}[c]{@{}l@{}}A car is described as follows: The buying price of this car is low. The maintenance cost for this car is low. The car can 2 people. The luggage boot in this \\ car is small. The safety rating of this car is estimated to be low. Is this car acceptable? Yes or No? Answer: Yes\\ \\ A car is described as follows: The buying price of this car is unknown The maintenance cost for this car is low. The car can 2 people. The luggage boot in \\ this car is small. The safety rating of this car is estimated to be high. Is this car acceptable? Yes or No? Answer: Yes\end{tabular} \\ \cmidrule(l){2-3} 
 & \multicolumn{1}{c|}{Query} & \begin{tabular}[c]{@{}l@{}}A car is described as follows: The buying price of this car is low. The maintenance cost for this car is low. The car can fit 4 or more people. The luggage\\ boot in this car is medium-sized or big. The safety rating of this car is estimated to be high. Is this car acceptable? Yes or No? Answer:\end{tabular} \\ \midrule
\multirow{2}{*}{Diabetes} & \multicolumn{1}{c|}{Demonstration} & \begin{tabular}[c]{@{}l@{}}The following describes the diagnostic measurements of a female patient of Pima Indian heritage. This patient is 34 or more years of age. She has been \\ pregnant 4 or more times. Her plasma glucose concentration at two hours in an oral glucose tolerance test is more than 121 milligrams per deciliter. Her \\ blood pressure is measured to be more than 69 mm Hg. She has a body mass index (BMI) of less than 32 kilograms per square meters and triceps skin \\ fold thickness of less than 21 mm. Her two-hours serum insulin is less than 80 microunits per milliliter. Her diabetes pedigree function is more than 0.5. \\ Does this patient have diabetes? Yes or No? Answer: No\\ \\ The following describes the diagnostic measurements of a female patient of Pima Indian heritage. This patient is less than 34 years of age. She has been \\ pregnant less than 4 times. Her plasma glucose concentration at two hours in an oral glucose tolerance test is less than 121 milligrams per deciliter. Her \\ blood pressure is measured to be less than 69 mm Hg. She has a body mass index (BMI) of more than 32 kilograms per square meters and triceps skin \\ fold thickness of less than 21 mm. Her two-hours serum insulin is less than 80 microunits per milliliter. Her diabetes pedigree function is more than 0.5. \\ Does this patient have diabetes? Yes or No? Answer: Yes\end{tabular} \\ \cmidrule(l){2-3} 
 & \multicolumn{1}{c|}{Query} & \begin{tabular}[c]{@{}l@{}}The following describes the diagnostic measurements of a female patient of Pima Indian heritage. This patient is less than 34 years of age. She has been \\ pregnant less than 4 times. Her plasma glucose concentration at two hours in an oral glucose tolerance test is less than 121 milligrams per deciliter. Her \\ blood pressure is measured to be less than 69 mm Hg. She has a body mass index (BMI) of less than 32 kilograms per square meters and triceps skin \\ fold thickness of less than 21 mm. Her two-hours serum insulin is less than 80 microunits per milliliter. Her diabetes pedigree function is less than 0.5. \\ Does this patient have diabetes? Yes or No? Answer:\end{tabular} \\ \midrule

\multirow{2}{*}{Heart} & \multicolumn{1}{c|}{Demonstration} &  \begin{tabular}[c]{@{}l@{}}The following describes diagnostic measurements of a patient. This patient is female and less than 54 years of age. She has angina chest pain. She has a \\resting blood pressure of more than 132 mm Hg. Her serum cholesterol is more than 199 milligrams per deciliter. Her fasting blood sugar is greater than \\120 milligrams per deciliter. Her resting electrocardiogram results show probable or definite left ventricular hypertrophy by Estes' criteria or have ST-T wave \\abnormality (T wave inversions and/or ST elevation or depression greater than 0.05 minute volume). The maximum heart rate achieved for this patient is \\less than 137. She has exercise induced angina. Her ST depression induced by exercise relative to rest is less than 0.9. The peak exercise ST segment for \\this patient has a slope. Does this patient have heart disease? Yes or No? Answer: No\\ \\The following describes diagnostic measurements of a patient. This patient is female and less than 54 years of age. She does not have angina chest pain. She \\has a resting blood pressure of less than 132 mm Hg. Her serum cholesterol is less than 199 milligrams per deciliter. Her fasting blood sugar is greater than 120 \\milligrams per deciliter. Her resting electrocardiogram results are normal. The maximum heart rate achieved for this patient is more than 137. She has exercise \\induced angina. Her ST depression induced by exercise relative to rest is more than 0.9. The peak exercise ST segment for this patient has a slope. Does this \\patient have heart disease? Yes or No? Answer: No\end{tabular}\\ \cmidrule(l){2-3} 
 & \multicolumn{1}{c|}{Query} &  \begin{tabular}[c]{@{}l@{}}The following describes diagnostic measurements of a patient. This patient is female and 54 or more years of age. She does not have angina chest pain. She has a \\resting blood pressure of less than 132 mm Hg. Her serum cholesterol is more than 199 milligrams per deciliter. Her fasting blood sugar is lower than 120 milligrams \\per deciliter. Her resting electrocardiogram results show probable or definite left ventricular hypertrophy by Estes' criteria or have ST-T wave abnormality (T wave \\inversions and/or ST elevation or depression greater than 0.05 minute volume). The maximum heart rate achieved for this patient is less than 137. She has exercise \\induced angina. er ST depression induced by exercise relative to rest is more than 0.9. The peak exercise ST segment for this patient is flat. Does this patient have\\ heart disease? Yes or No? Answer:\end{tabular}\\ \midrule
\multirow{2}{*}{Jungle} & \multicolumn{1}{c|}{Demonstration} & \begin{tabular}[c]{@{}l@{}}In a two-piece endgame of jungle chess, the white piece has strength less than 4.2 and is on file 3 or more and rank less than 4. The black piece has strength \\ less than 4.2 and is on file 3 or more and rank less than 4. Does the white player win this game? Yes or No? Answer: No\\ \\ In a two-piece endgame of jungle chess, the white piece has strength more than 4.2 and is on file less than 3 and rank 3 or more. The black piece has strength \\ less than 4.2 and is on file 3 or more and rank less than 4. Does the white player win this game? Yes or No? Answer: Yes\end{tabular} \\ \cmidrule(l){2-3} 
 & \multicolumn{1}{c|}{Query} & \begin{tabular}[c]{@{}l@{}}In a two-piece endgame of jungle chess, the white piece has strength less than 4.2 and is on file 3 or more and rank 3 or more. The black piece has strength \\ less than 4.2 and is on file less than 3 and rank 3 or more. Does the white player win this game? Yes or No? Answer:\end{tabular} \\ \bottomrule
\end{tabular}%
}
\end{table*}

\paragraph{\textbf{GDP}}
\label{app:gdp-examples}
In Table \ref{tab:prompt-examples} we show prompt examples generated for the GDP-TabICL settings for each dataset. As opposed to LDP-TabICL, since binarization was not performed in GDP-TabICL, phrases in the description and query can be concrete (e.g., ``39 year old male"). 
\begin{table*}[h!]
\centering
\caption{Example prompts for GDP-TabICL when $k=2$ and $\epsilon=5$.}
\label{tab:prompt-examples}
\resizebox{\textwidth}{!}{%
\begin{tabular}{@{}c|cl@{}}
\toprule
Dataset & \multicolumn{2}{c}{Prompt} \\ \midrule
\multirow{2}{*}{Adult} & \multicolumn{1}{c|}{Demonstration} & \begin{tabular}[c]{@{}l@{}}An individual recorded in the 1994 US census is described as follows: This person is a 42 years old male. He has a Bachelor's degree at most. \\ He works in the private sector. His occupation is in executive management. He works 89 hours per week. He had a capital gain of 3948.34 and a \\ capital loss of 90.96 last year. He is from United-States. He is White. He is married to a civilian. He is the husband of the other person in his household. \\ Does this person earn more than 50,000 dollars annually? Yes or No? Answer: Yes\\ \\ An individual recorded in the 1994 US census is described as follows: This person is a 37 years old male. He has a high school degree at most. He works \\ in the private sector. His occupation is in craft and/or repair. He works 40 hours per week. He had a capital gain of 141.83 and a capital loss of 245.1 last \\ year. He is from United-States. He is White. He has never been married. He is the husband of the other person in his household. Does this person earn more \\ than 50,000 dollars annually? Yes or No? Answer: No\end{tabular} \\ \cmidrule(l){2-3} 
 & \multicolumn{1}{c|}{Query} & \begin{tabular}[c]{@{}l@{}}An individual recorded in the 1994 US census is described as follows: This person is a 25 years old male. He has completed eleventh grade at most. \\ He works in the private sector. His occupation is in machine operation and inspection. He works 40 hours per week. He had a capital gain of 0 and a \\ capital loss of 0 last year. He is from United-States. He is Black. He has never been married. He is the child of the other person in his household. Does this \\ person earn more than 50,000 dollars annually? Yes or No?\end{tabular} \\ \midrule
\multirow{2}{*}{Bank} & \multicolumn{1}{c|}{Demonstration} & \begin{tabular}[c]{@{}l@{}}A client at a Portuguese banking institution is described as follows: The client is 38 years old. They have attained secondary-level education. Their job is \\ in management. They are married. Their average yearly balance is -817.14 euros. They do not have credit in default. They do not have any housing loans. \\ They do not have any personal loans. The contact communication type is cellular. They were contacted 0 times during this campaign. They were last contacted\\ on May 16 for a duration of 474 seconds. They were contacted 0 times in a previous campaign and were last contacted 59 days ago for the previous campaign.\\ The outcome of the previous marketing campaign is not known for this client. Does this person subscribe to a term deposit? Yes or No? Answer: Yes\\ \\ A client at a Portuguese banking institution is described as follows: The client is 40 years old. They have attained secondary-level education. They have a blue-collar \\ job. They are married. Their average yearly balance is 1457.56 euros. They do not have credit in default. They have housing loans. They do not have any personal \\ loans. The contact communication type is cellular. They were contacted 3 times during this campaign. They were last contacted on May 16 for a duration of 217 \\ seconds. They were contacted 0 times in a previous campaign and were last contacted 37 days ago for the previous campaign. The outcome of the previous marketing\\ campaign is not known for this client. Does this person subscribe to a term deposit? Yes or No? Answer: No\end{tabular} \\ \cmidrule(l){2-3} 
 & \multicolumn{1}{c|}{Query} & \begin{tabular}[c]{@{}l@{}}A client at a Portuguese banking institution is described as follows: The client is 30 years old. They have attained secondary-level education. Their job is a \\ technician. They are single. Their average yearly balance is 383.0 euros. They do not have credit in default. They have housing loans. They have personal loans. \\ The contact communication type is cellular. They were contacted 3 times during this campaign. They were last contacted on November 19 for a duration of 145 \\ seconds. They were not contacted in previous campaigns. The outcome of the previous marketing campaign is not known for this client. Does this person subscribe \\ to a term deposit? Yes or No?\end{tabular} \\ \midrule
\multirow{2}{*}{Blood} & \multicolumn{1}{c|}{Demonstration} & \begin{tabular}[c]{@{}l@{}}A blood donor at a Blood Transfusion Service Center is described as follows: The donor has donated blood 7 times. They have donated a total of 2062.93 c.c. of blood. \\ They last donated blood 4 months ago. Their first blood donation was 37 months ago. Did the donor donate blood in March 2007? Yes or No? Answer: Yes\\ \\ A blood donor at a Blood Transfusion Service Center is described as follows: The donor has donated blood 4 times. They have donated a total of 291.97 c.c. of blood. \\ They last donated blood 11 months ago. Their first blood donation was 35 months ago. Did the donor donate blood in March 2007? Yes or No? Answer: No\end{tabular} \\ \cmidrule(l){2-3} 
 & \multicolumn{1}{c|}{Query} & \begin{tabular}[c]{@{}l@{}}A blood donor at a Blood Transfusion Service Center is described as follows: The donor has donated blood 2 times. They have donated a total of 500.0 c.c. of blood. \\ They last donated blood 11 months ago. Their first blood donation was 21 months ago. Did the donor donate blood in March 2007? Yes or No?\end{tabular} \\ \midrule
\multirow{2}{*}{Calhousing} & \multicolumn{1}{c|}{Demonstration} & \begin{tabular}[c]{@{}l@{}}A house block in California has the following attributes according to the 1990 California census. This housing block is located at latitude 35.2 and longitude -119.46. \\ The houses in the block have a total of 2896 rooms with 553 bedrooms. The median age of houses in the block is 30 years. There are a total of 530 households in the \\ block with a total population of 1431. The median income of the households in the block is 1681 thousand dollars. Is this housing block valuable? Yes or No? Answer:Yes\\   \\ A house block in California has the following attributes according to the 1990 California census. This housing block is located at latitude 35.86 and longitude -120.36. \\ The houses in the block have a total of 2377 rooms with 521 bedrooms. The median age of houses in the block is 28 years. There are a total of 473 households in the \\ block with a total population of 1454. The median income of the households in the block is 3130 thousand dollars. Is this housing block valuable? Yes or No? Answer: No\end{tabular} \\ \cmidrule(l){2-3} 
 & \multicolumn{1}{c|}{Query} & \begin{tabular}[c]{@{}l@{}}A house block in California has the following attributes according to the 1990 California census. This housing block is located at latitude 33.72 and longitude -117.99. \\ The houses in the block have a total of 1787 rooms with 275 bedrooms. The median age of houses in the block is 26 years. There are a total of 270 households in the block \\ with a total population of 801. The median income of the households in the block is 56 thousand dollars. Is this housing block valuable? Yes or No?\end{tabular} \\ \midrule
\multirow{2}{*}{Car} & \multicolumn{1}{c|}{Demonstration} & \begin{tabular}[c]{@{}l@{}}A car is described as follows: The buying price of this car is very high. The maintenance cost for this car is very high. The car has 2 doors. The car can fit 2 people. The \\ luggage boot in this car is small. The safety rating of this car is estimated to be low. Is this car acceptable? Yes or No? Answer: No\\ \\ A car is described as follows: The buying price of this car is low. The maintenance cost for this car is medium. The car has 3 doors. The car can fit 4 people. The luggage \\ boot in this car is big. The safety rating of this car is estimated to be high. Is this car acceptable? Yes or No? Answer: Yes\end{tabular} \\ \cmidrule(l){2-3} 
 & \multicolumn{1}{c|}{Query} & \begin{tabular}[c]{@{}l@{}}A car is described as follows: The buying price of this car is high. The maintenance cost for this car is low. The car has 3 doors. The car can fit 4 people. The luggage \\ boot in this car is small. The safety rating of this car is estimated to be high. Is this car acceptable? Yes or No?\end{tabular} \\ \midrule
\multirow{2}{*}{Diabetes} & \multicolumn{1}{c|}{Demonstration} & \begin{tabular}[c]{@{}l@{}}The following describes the diagnostic measurements of a female patient of Pima Indian heritage. This patient is 37 years old. She has been pregnant 5 times. Her plasma \\ glucose concentration at two hours in an oral glucose tolerance test is 127.45 milligrams per deciliter. Her blood pressure is measured to be 70.11 mm Hg. She has a body \\ mass index (BMI) of 106.39 kilograms per square meters and triceps skin fold thickness of 28.61 mm. Her two-hours serum insulin is 80.41 microunits per milliliter. Her \\ diabetes pedigree function is 2.42. Does this patient have diabetes? Yes or No? Answer: Yes\\ \\ The following describes the diagnostic measurements of a female patient of Pima Indian heritage. This patient is 29 years old. She has been pregnant 3 times. Her plasma \\ glucose concentration at two hours in an oral glucose tolerance test is 113.93 milligrams per deciliter. Her blood pressure is measured to be 71.78 mm Hg. She has a body\\ mass index (BMI) of 26.83 kilograms per square meters and triceps skin fold thickness of 18.95 mm. Her two-hours serum insulin is 55.76 microunits per milliliter. Her \\ diabetes pedigree function is 1.68. Does this patient have diabetes? Yes or No? Answer: No\end{tabular} \\ \cmidrule(l){2-3} 
 & \multicolumn{1}{c|}{Query} & \begin{tabular}[c]{@{}l@{}}The following describes the diagnostic measurements of a female patient of Pima Indian heritage. This patient is 34 years old. She has been pregnant 5 times. Her plasma\\  glucose concentration at two hours in an oral glucose tolerance test is 155.0 milligrams per deciliter. Her blood pressure is measured to be 84.0 mm Hg. She has a body\\  mass index (BMI) of 38.7 kilograms per square meters and triceps skin fold thickness of 44.0 mm. Her two-hours serum insulin is 545.0 microunits per milliliter. Her\\  diabetes pedigree function is 0.619. Does this patient have diabetes? Yes or No?\end{tabular} \\ \midrule
\multirow{2}{*}{Heart} & \multicolumn{1}{c|}{Demonstration} & \begin{tabular}[c]{@{}l@{}}The following describes diagnostic measurements of a patient. This patient is a 57 years old male. He has asymptomatic chest pain. He has a resting blood pressure of \\ 135.26 mm Hg. His serum cholesterol is 204.1 milligrams per deciliter. He has fasting blood sugar lower than 120 milligrams per decilitre. His resting electrocardiogram \\ results are normal. The maximum heart rate achieved for this patient is 129.21. He has exercise induced angina. His ST depression induced by exercise relative to rest is \\ 1.33. The peak exercise ST segment for this patient is flat. Does this patient have heart disease? Yes or No? Answer: Yes\\ \\ The following describes diagnostic measurements of a patient. This patient is a 53 years old male. He has atypical angina chest pain. He has a resting blood pressure of \\ 138.03 mm Hg. His serum cholesterol is 232.2 milligrams per deciliter. He has fasting blood sugar lower than 120 milligrams per decilitre. His resting electrocardiogram\\ results are normal. The maximum heart rate achieved for this patient is 164.16. He does not have exercise induced angina. His ST depression induced by exercise relative\\ to rest is 0.61. The peak exercise ST segment for this patient slopes downwards. Does this patient have heart disease? Yes or No? Answer: No\end{tabular} \\ \cmidrule(l){2-3} 
 & \multicolumn{1}{c|}{Query} & \begin{tabular}[c]{@{}l@{}}The following describes diagnostic measurements of a patient. This patient is a 50 years old female. She has atypical angina chest pain. She has a resting blood pressure of \\ 110 mm Hg. Her serum cholesterol is 202 milligrams per deciliter. She has fasting blood sugar lower than 120 milligrams per decilitre. Her resting electrocardiogram results \\ are normal. The maximum heart rate achieved for this patient is 145. She does not have exercise induced angina. Her ST depression induced by exercise relative to rest is 0.0.\\ The peak exercise ST segment for this patient slopes downwards. Does this patient have heart disease? Yes or No?\end{tabular} \\ \midrule
\multirow{2}{*}{Jungle} & \multicolumn{1}{c|}{Demonstration} & \begin{tabular}[c]{@{}l@{}}In a two-piece endgame of jungle chess, the white piece has strength 4 and is on file 3 and rank 5. The black piece has strength 4 and is on file 3 and rank 5. Does the white \\ player win this game? Yes or No? Answer: Yes\\ \\ In a two-piece endgame of jungle chess, the white piece has strength 4 and is on file 3 and rank 3. The black piece has strength 5 and is on file 3 and rank 3. Does the white\\  player win this game? Yes or No? Answer: No\end{tabular} \\ \cmidrule(l){2-3} 
 & \multicolumn{1}{c|}{Query} & \begin{tabular}[c]{@{}l@{}}In a two-piece endgame of jungle chess, the white piece has strength 0 and is on file 2 and rank 2. The black piece has strength 6 and is on file 3 and rank 4. Does the white\\  player win this game? Yes or No?\end{tabular} \\ \bottomrule
\end{tabular}%
}
\end{table*}

\subsection{Additional Experimental Results}

\subsubsection{Computational Cost}
\label{app:comp-cost}
All experiments in this work were run on a Tesla V100 GPU (32GB RAM) which has a maximum power consumption of 300W. The carbon efficiency of the local power grid is 0.3633 kg/kWH.\footnote{\url{https://www.epa.gov/egrid/data-explorer}} Based on these two values, we calculated the average kg CO$_2$eq values where 
\begin{equation}
\label{eq:comp-cost}\small
    \textrm{kg CO}_2\textrm{eq} = \textrm{Power (W)} \times \textrm{Time (H)} / 1000 * \textrm{kg/kWH}
\end{equation}
Table \ref{tab:comp-cost} details the average time to query Llama-2-13B for every $\epsilon$ value [1, 5, 10, 25, 50] on a single $k$ setting. In other words, Table \ref{tab:comp-cost} show the average time and CO$_2$eq cost to run our GDP experiments once. We note that we run all of our experiments five times, meaning the true time/CO$_2$eq cost is five times that reported in the table. In Table \ref{tab:per-run-cost} we report the time and CO$_2$eq cost to query Llama-2-13b on a single prompt in the GDP-TabICL setting. We note that the value of $\epsilon$ had marginal effect on the running time, and therefore we did not report the values per $\epsilon$, $k$ combination but rather only per $k$. Additionally, we only report results for the GDP case as all timing results for the LDP case were equivalent or below the values reported in Tables \ref{tab:comp-cost} and \ref{tab:per-run-cost}. Estimations were conducted using the Machine Learning Impact calculator\footnote{\url{https://mlco2.github.io/impact#compute}} presented in \cite{lacoste2019quantifying}.

\begin{table}[H]
\centering
\caption{Average time in hours and kilogram CO$_2$eq. (hrs/kg) spent for running experiments for every $\epsilon$ value [1, 2, 5, 10, 25, 50] on a single $k$ for GDP-TabICL on Llama-2-13B.}
\label{tab:comp-cost}
\resizebox{\columnwidth}{!}{%
\begin{tabular}{@{}ccccccccc@{}}
\toprule
 & \multicolumn{8}{c}{\textbf{Dataset}} \\
$\bm{k}$ & \textbf{Adult} & \textbf{Bank} & \textbf{Blood} & \textbf{Calhousing} & \textbf{Car} & \textbf{Diabetes} & \textbf{Heart} & \textbf{Jungle} \\ \midrule
1 & 0.45 / 0.049 & 0.29 / 0.032 & 0.05 / 0.005 & 0.14 / 0.015 & 0.08 / 0.009 & 0.06 / 0.007 & 0.07 / 0.007 & 0.21 / 0.023 \\
2 & 0.68 / 0.074 & 0.43 / 0.047 & 0.06 / 0.006 & 0.19 / 0.020 & 0.08 / 0.009 & 0.08 / 0.009 & 0.09 / 0.009 & 0.23 / 0.025 \\
4 & 1.02 / 0.111 & 0.74 / 0.080 & 0.07 / 0.008 & 0.29 / 0.031 & 0.13 / 0.014 & 0.12 / 0.013 & 0.13 / 0.014 & 0.30 / 0.032 \\
8 & 1.93 / 0.210 & 1.43 / 0.156 & 0.11 / 0.012 & 0.47 / 0.051 & 0.19 / 0.021 & 0.19 / 0.021 & 0.24 / 0.026 & 0.45 / 0.049 \\ \bottomrule
\end{tabular}%
}
\end{table}

\begin{table}[H]
\centering
\caption{Average time in seconds and gram CO$_2$eq. (s/g) spent for running a single prompt on one $\epsilon$ value per $k$ for GDP-TabICL. Time results equal Table \ref{tab:comp-cost} value * 3600 / 6 / $N_{test}$. To get g CO$_2$\textsubscript{eq}, convert the calculated seconds to hours, and multiply the result of Eq. \ref{eq:comp-cost} by 1000.}
\label{tab:per-run-cost}
\resizebox{\columnwidth}{!}{%
\begin{tabular}{@{}ccccccccc@{}}
\toprule
 & \multicolumn{8}{c}{\textbf{Dataset}} \\
$\bm{k}$ & \textbf{Adult} & \textbf{Bank} & \textbf{Blood} & \textbf{Calhousing} & \textbf{Car} & \textbf{Diabetes} & \textbf{Heart} & \textbf{Jungle} \\ \midrule
1 & 0.18 / 0.005 & 0.20 / 0.006 & 0.26 / 0.008 & 0.20 / 0.006 & 0.19 / 0.006 & 0.32 / 0.010 & 0.29 / 0.009 & 0.14 / 0.004\\
2 & 0.27 / 0.008 & 0.29 / 0.009 & 0.30 / 0.009 & 0.27 / 0.008 & 0.20 / 0.006 & 0.41 / 0.012 & 0.37 / 0.012 & 0.15 / 0.005\\
4 & 0.41 / 0.012 & 0.49 / 0.015 & 0.39 / 0.071 & 0.42 / 0.013 & 0.29 / 0.009 & 0.64 / 0.019 & 0.56 / 0.017 & 0.20 / 0.006\\
8 & 0.77 / 0.023 & 0.95 / 0.029 & 0.59 / 1.064 & 0.68 / 1.237 & 0.45 / 0.014 & 1.11 / 0.034 & 1.03 / 0.031 & 0.30 / 0.009\\ \bottomrule
\end{tabular}%
}
\end{table}
\subsubsection{Results on Llama-2-7B}
\label{app:7B}
\paragraph{\textbf{LDP-TabICL}}
We report the average accuracy and standard deviation over five runs of LDP-TabICL on all datasets and combinations of $k$ and $\epsilon$ on the Llama-2-7B model in Table \ref{tab:gdp-tabicl-results}. Similar to the results on the Llama-2-13B model, there is no clear trend between $\epsilon$ increasing and the accuracy changing. However, as opposed to Llama-2-13B, on the Llama-2-7B model there is also no trend between accuracy and the $k$ value when the balanced and unbalanced datasets are analyzed separately. In general, however, the results on the Llama-2-7B model, while showing weaker trends than the Llama-2-13B model, achieve comparable performance to the Llama-2-13B model. This result is further supported by the $p$-values presented in Table \ref{tab:gdp-pvalue} of Section \ref{sec:eval}.

\begin{table}[H]
\centering
\caption{LDP-TabICL average accuracy and standard deviation over 5 runs on Llama-2-7B using differing number of splits $k$ and privacy budget value $\epsilon$. LR: logistic regression, GNB: Gaussian Na\"ive Bayes, *: model trained on reconstructed LDP data, \underline{Datset}: unbalanced dataset.}
\label{tab:ldp-tabicl-results}
\resizebox{\columnwidth}{!}{%
\begin{tabular}{@{}ccccHccccc@{}}
\toprule
 & \textbf{} & \multicolumn{8}{c}{$\bm{\epsilon}$} \\
\textbf{Dataset} & \textbf{Method} & $\bm{k}$ & \textbf{1} & \textbf{2} & \textbf{5} & \textbf{10} & \textbf{25} & \textbf{50} & $\bm{\infty}$\\ \midrule
\multirow{6}{*}{\underline{Adult}} & LDP-TabICL & 1 & 0.75\textsubscript{.05} & 0.72\textsubscript{.08} & 0.74\textsubscript{.07} & 0.77\textsubscript{.03} & 0.78\textsubscript{.06} & 0.80\textsubscript{.02} & 0.73\textsubscript{.00} \\
                       & LDP-TabICL & 2 & 0.73\textsubscript{.08} & 0.76\textsubscript{.04} & 0.77\textsubscript{.04} & 0.73\textsubscript{.13} & 0.78\textsubscript{.01} & 0.78\textsubscript{.02} & 0.75\textsubscript{.00} \\
                       & LDP-TabICL & 4 & 0.75\textsubscript{.04} & \textbf{0.78\textsubscript{.01}} & 0.77\textsubscript{.03} & \textbf{0.79\textsubscript{.03}} & 0.78\textsubscript{.01} & 0.77\textsubscript{.02} & 0.73\textsubscript{.00} \\
                       & LDP-TabICL & 8 & \textbf{0.79\textsubscript{.01}} & 0.77\textsubscript{.01} & \textbf{0.79\textsubscript{.02}} & 0.78\textsubscript{.01} & 0.78\textsubscript{.01} & 0.76\textsubscript{.01} & 0.77\textsubscript{.00} \\
                       & LR*         & - & 0.35\textsubscript{.20} & 0.59\textsubscript{.12} & 0.54\textsubscript{.16} & 0.70\textsubscript{.06} & \textbf{0.82\textsubscript{.00}} & \textbf{0.83\textsubscript{.00}} & \textbf{0.85\textsubscript{.00}} \\
                      & GNB*         & - & 0.38\textsubscript{.20} & 0.60\textsubscript{.13} & 0.57\textsubscript{.15} & 0.71\textsubscript{.06} & \textbf{0.82\textsubscript{.00}} & 0.81\textsubscript{.00} & 0.81\textsubscript{.00} \\\midrule
\multirow{6}{*}{\underline{Bank}} & LDP-TabICL& 1 & 0.61\textsubscript{.32} & 0.76\textsubscript{.08} & 0.78\textsubscript{.19} & 0.67\textsubscript{.29} & 0.73\textsubscript{.09} & 0.84\textsubscript{.04} & 0.77\textsubscript{.00} \\
                      & LDP-TabICL& 2 & \textbf{0.88\textsubscript{.01}} & \textbf{0.86\textsubscript{.05}} & \textbf{0.88\textsubscript{.01}} & 0.79\textsubscript{.17} & 0.88\textsubscript{.01} & 0.88\textsubscript{.02} & 0.88\textsubscript{.00} \\
                      & LDP-TabICL& 4 & 0.80\textsubscript{.08} & 0.84\textsubscript{.09} & \textbf{0.88\textsubscript{.01}} & \textbf{0.89\textsubscript{.01}} & 0.88\textsubscript{.01} & 0.87\textsubscript{.02} & 0.87\textsubscript{.00} \\
                      & LDP-TabICL& 8 & 0.55\textsubscript{.28} & 0.73\textsubscript{.34} & 0.67\textsubscript{.31} & 0.86\textsubscript{.03} & 0.86\textsubscript{.04} & 0.88\textsubscript{.02} & 0.84\textsubscript{.00} \\
                      & LR*        & - & 0.57\textsubscript{.37} & 0.45\textsubscript{.35} & 0.67\textsubscript{.14} & 0.88\textsubscript{.01} & \textbf{0.89\textsubscript{.00}} & \textbf{0.89\textsubscript{.00}} & \textbf{0.90\textsubscript{.00}} \\
 & GNB*         & - & 0.58\textsubscript{.37} & 0.44\textsubscript{.35} & 0.69\textsubscript{.13} & 0.88\textsubscript{.01} & 0.88\textsubscript{.00} & 0.88\textsubscript{.00} & 0.88\textsubscript{.00} \\\midrule
 
 \multirow{6}{*}{\underline{Blood}} & LDP-TabICL& 1 & 0.67\textsubscript{.20} & 0.55\textsubscript{.19} & 0.76\textsubscript{.03} & 0.58\textsubscript{.19} & 0.75\textsubscript{.05} & 0.68\textsubscript{.13} & 0.74\textsubscript{.00} \\
                        & LDP-TabICL& 2 & 0.77\textsubscript{.02} & \textbf{0.79\textsubscript{.01}} & \textbf{0.78\textsubscript{.02}} & \textbf{0.77\textsubscript{.03}} & \textbf{0.78\textsubscript{.02}} & \textbf{0.78\textsubscript{.01}} & \textbf{0.78\textsubscript{.00}} \\
                        & LDP-TabICL& 4 & 0.77\textsubscript{.05} & 0.75\textsubscript{.07} & 0.74\textsubscript{.06} & \textbf{0.77\textsubscript{.03}} & \textbf{0.78\textsubscript{.01}} & 0.74\textsubscript{.05} & 0.75\textsubscript{.00} \\
                        & LDP-TabICL& 8 & \textbf{0.78\textsubscript{.02}} & 0.77\textsubscript{.03} & \textbf{0.78\textsubscript{.04}} & 0.57\textsubscript{.17} & 0.74\textsubscript{.03} & 0.76\textsubscript{.06} & 0.73\textsubscript{.00} \\
                        & LR*        & - & 0.51\textsubscript{.22} & 0.52\textsubscript{.21} & 0.74\textsubscript{.04} & \textbf{0.77\textsubscript{.02}} & \textbf{0.78\textsubscript{.01}} & 0.77\textsubscript{.04} & 0.74\textsubscript{.00} \\
 & GNB*         & - & 0.48\textsubscript{.20} & 0.53\textsubscript{.20} & 0.75\textsubscript{.04} & 0.71\textsubscript{.06} & 0.73\textsubscript{.03} & 0.77\textsubscript{.02} & 0.74\textsubscript{.03} \\\midrule

 \multirow{6}{*}{\underline{Car}} & LDP-TabICL& 1 & 0.68\textsubscript{.04} & 0.70\textsubscript{.03} & 0.61\textsubscript{.16} & 0.53\textsubscript{.14} & 0.63\textsubscript{.14} & 0.53\textsubscript{.08} & 0.68\textsubscript{.00} \\
                      & LDP-TabICL& 2 & \textbf{0.74\textsubscript{.04}} & 0.71\textsubscript{.03} & 0.70\textsubscript{.03} & 0.72\textsubscript{.02} & 0.71\textsubscript{.02} & 0.72\textsubscript{.02} & 0.72\textsubscript{.00} \\
                      & LDP-TabICL& 4 & \textbf{0.74\textsubscript{.02}} & \textbf{0.74\textsubscript{.02}} & \textbf{0.74\textsubscript{.02}} & 0.73\textsubscript{.02} & 0.76\textsubscript{.05} & 0.71\textsubscript{.02} & 0.73\textsubscript{.00} \\
                      & LDP-TabICL& 8 & \textbf{0.74\textsubscript{.04}} & 0.71\textsubscript{.02} & 0.73\textsubscript{.02} & \textbf{0.74\textsubscript{.04}} & 0.70\textsubscript{.02} & 0.72\textsubscript{.04} & 0.71\textsubscript{.00} \\
                      & LR*        & - & 0.53\textsubscript{.18} & 0.38\textsubscript{.06} & 0.57\textsubscript{.14} & 0.62\textsubscript{.07} & \textbf{0.90\textsubscript{.02}} & \textbf{0.94\textsubscript{.01}} & \textbf{0.95\textsubscript{.00}} \\
  & GNB*         & - & 0.50\textsubscript{.01} & 0.54\textsubscript{.07} & 0.54\textsubscript{.14} & 0.63\textsubscript{.06} & 0.74\textsubscript{.01} & 0.73\textsubscript{.01} & 0.73\textsubscript{.01} \\\midrule
  
 \multirow{6}{*}{Calhousing} & LDP-TabICL& 1 & \textbf{0.59\textsubscript{.06}} & \textbf{0.60\textsubscript{.04}} & \textbf{0.56\textsubscript{.04}} & 0.53\textsubscript{.09} & 0.57\textsubscript{.05} & 0.53\textsubscript{.05} & 0.57\textsubscript{.00} \\
                             & LDP-TabICL& 2 & 0.50\textsubscript{.03} & 0.51\textsubscript{.06} & 0.51\textsubscript{.04} & 0.54\textsubscript{.07} & 0.54\textsubscript{.07} & 0.52\textsubscript{.09} & 0.52\textsubscript{.00} \\
                             & LDP-TabICL& 4 & 0.52\textsubscript{.04} & 0.52\textsubscript{.04} & 0.49\textsubscript{.03} & 0.50\textsubscript{.06} & 0.49\textsubscript{.02} & 0.52\textsubscript{.07} & 0.55\textsubscript{.00} \\
                             & LDP-TabICL& 8 & 0.53\textsubscript{.07} & 0.55\textsubscript{.08} & 0.55\textsubscript{.03} & 0.51\textsubscript{.02} & 0.52\textsubscript{.07} & 0.56\textsubscript{.09} & 0.54\textsubscript{.00} \\
                             & LR*        & - & 0.50\textsubscript{.01} & 0.54\textsubscript{.07} & 0.52\textsubscript{.14} & \textbf{0.63\textsubscript{.07}} & 0.75\textsubscript{.00} & 0.75\textsubscript{.00} & 0.84\textsubscript{.00} \\
 & GNB*         & - & 0.52\textsubscript{.17} & 0.37\textsubscript{.05} & \textbf{0.56\textsubscript{.15}} & 0.61\textsubscript{.07} & \textbf{0.91\textsubscript{.01}} & \textbf{0.88\textsubscript{.01}} & \textbf{0.85\textsubscript{.01}} \\\midrule

 \multirow{6}{*}{Diabetes} & LDP-TabICL& 1 & 0.67\textsubscript{.02} & \textbf{0.67\textsubscript{.04}} & \textbf{0.68\textsubscript{.03}} & \textbf{0.70\textsubscript{.01}} & 0.68\textsubscript{.01} & 0.70\textsubscript{.02} & 0.66\textsubscript{.00} \\
                           & LDP-TabICL& 2 & 0.66\textsubscript{.04} & 0.63\textsubscript{.04} & 0.65\textsubscript{.02} & 0.64\textsubscript{.01} & 0.65\textsubscript{.03} & 0.65\textsubscript{.03} & 0.65\textsubscript{.00} \\
                           & LDP-TabICL& 4 & 0.56\textsubscript{.12} & 0.60\textsubscript{.08} & 0.66\textsubscript{.03} & 0.66\textsubscript{.04} & 0.67\textsubscript{.04} & 0.64\textsubscript{.01} & 0.64\textsubscript{.00} \\
                           & LDP-TabICL& 8 & \textbf{0.68\textsubscript{.02}} & 0.56\textsubscript{.11} & 0.56\textsubscript{.18} & 0.60\textsubscript{.14} & 0.58\textsubscript{.13} & 0.65\textsubscript{.02} & 0.64\textsubscript{.00} \\
                           & LR*        & - & 0.57\textsubscript{.11} & 0.48\textsubscript{.16} & 0.48\textsubscript{.17} & 0.59\textsubscript{.07} & \textbf{0.72\textsubscript{.02}} & \textbf{0.72\textsubscript{.02}} & \textbf{0.79\textsubscript{.00}} \\
  & GNB*         & - & 0.58\textsubscript{.12} & 0.50\textsubscript{.15} & 0.47\textsubscript{.16} & 0.60\textsubscript{.06} & 0.70\textsubscript{.02} & 0.71\textsubscript{.02} & 0.71\textsubscript{.03} \\\midrule
  
 \multirow{6}{*}{Heart} & LDP-TabICL & 1 & 0.56\textsubscript{.10} & 0.47\textsubscript{.08} & \textbf{0.63\textsubscript{.09}} & 0.52\textsubscript{.05} & 0.54\textsubscript{.11} & 0.64\textsubscript{.11} & 0.57\textsubscript{.00} \\
                        & LDP-TabICL & 2 & \textbf{0.57\textsubscript{.07}} & 0.44\textsubscript{.04} & 0.44\textsubscript{.02} & 0.50\textsubscript{.08} & 0.47\textsubscript{.07} & 0.54\textsubscript{.10} & 0.50\textsubscript{.00} \\
                        & LDP-TabICL & 4 & 0.50\textsubscript{.10} & 0.49\textsubscript{.11} & 0.59\textsubscript{.12} & 0.47\textsubscript{.08} & 0.58\textsubscript{.14} & 0.53\textsubscript{.10} & 0.50\textsubscript{.00} \\
                        & LDP-TabICL & 8 & 0.46\textsubscript{.12} & \textbf{0.54\textsubscript{.06}} & 0.47\textsubscript{.06} & 0.55\textsubscript{.10} & 0.64\textsubscript{.01} & 0.61\textsubscript{.02} & 0.64\textsubscript{.00} \\
                        & LR*         & - & 0.46\textsubscript{.16} & 0.52\textsubscript{.16} & 0.59\textsubscript{.09} & \textbf{0.56\textsubscript{.17}} & 0.67\textsubscript{.07} & 0.78\textsubscript{.01} & \textbf{0.85\textsubscript{.00}} \\
 & GNB*         & - & 0.46\textsubscript{.16} & 0.52\textsubscript{.15} & 0.60\textsubscript{.09} & \textbf{0.56\textsubscript{.16}} & \textbf{0.71\textsubscript{.05}} & \textbf{0.79\textsubscript{.03}} & \textbf{0.85\textsubscript{.01}} \\\midrule
 
 \multirow{6}{*}{Jungle} & LDP-TabICL & 1 & 0.50\textsubscript{.02} & \textbf{0.51\textsubscript{.02}} & 0.51\textsubscript{.03} & 0.52\textsubscript{.03} & 0.52\textsubscript{.03} & 0.48\textsubscript{.00} & 0.51\textsubscript{.00} \\
                         & LDP-TabICL & 2 & 0.49\textsubscript{.01} & 0.50\textsubscript{.02} & 0.49\textsubscript{.02} & 0.47\textsubscript{.05} & 0.50\textsubscript{.01} & 0.48\textsubscript{.02} & 0.50\textsubscript{.00} \\
                         & LDP-TabICL & 4 & 0.50\textsubscript{.03} & \textbf{0.51\textsubscript{.02}} & 0.50\textsubscript{.03} & 0.50\textsubscript{.00} & 0.48\textsubscript{.03} & 0.51\textsubscript{.01} & 0.50\textsubscript{.00} \\
                         & LDP-TabICL & 8 & 0.52\textsubscript{.04} & 0.49\textsubscript{.01} & \textbf{0.52\textsubscript{.03}} & 0.50\textsubscript{.03} & 0.52\textsubscript{.03} & 0.51\textsubscript{.03} & 0.51\textsubscript{.00} \\
                         & LR*         & - & \textbf{0.56\textsubscript{.07}} & 0.47\textsubscript{.05} & 0.48\textsubscript{.08} & 0.73\textsubscript{.00} & 0.73\textsubscript{.00} & 0.73\textsubscript{.00} & 0.73\textsubscript{.00} \\
                          & GNB*         & - & \textbf{0.56\textsubscript{.07}} & 0.47\textsubscript{.04} & 0.47\textsubscript{.08} & \textbf{0.74\textsubscript{.00}} & \textbf{0.74\textsubscript{.00}} & \textbf{0.74\textsubscript{.00}} & \textbf{0.74\textsubscript{.00}} \\\bottomrule
\end{tabular}%
}
\end{table}

\paragraph{\textbf{GDP-TabICL}}
We present the results for inference on Llama-2-7B using GDP-TabICL on all datasets in Table ~\ref{tab:gdp-tabicl-results}. Unlike the GDP-TabICL results on the Llama-2-13B model reported in Section \ref{sec:eval}, no real trend exists between the accuracy and $\epsilon$ increasing, even when the datasets are analyzed under the lens of if they are balanced or unbalanced. Further, there is no trend of the unbalanced datasets performing best with few demonstration examples while balanced datasets perform best with more demonstration examples. Rather, for most datasets, using more demonstration examples produced better results in general. More interestingly, when comparing the Llama-2-7B and Llama-2-13B results, the 7B model performs better on the unbalanced datasets while the 13B model performs better on the balanced. Future research will look into why this phenomenon occurs. 

\begin{table}[H]
\centering
\caption{GDP-TabICL average accuracy and standard deviation over 5 runs on Llama-2-7B using differing number of splits $k$ and privacy budget value $\epsilon$. LR: logistic regression, GNB: Gaussian Na\"ive Bayes, \underline{Dataset}: unbalanced dataset.}
\label{tab:gdp-tabicl-results}
\resizebox{\columnwidth}{!}{%
\begin{tabular}{@{}ccccHccccc@{}}
\toprule
 & \textbf{} & \multicolumn{7}{c}{$\bm{\epsilon}$} \\
\textbf{Dataset} & \textbf{Method} & $\bm{k}$ & \textbf{1} & \textbf{2} & \textbf{5} & \textbf{10} & \textbf{25} & \textbf{50} & $\bm{\infty}$\\ \midrule
\multirow{6}{*}{\underline{Adult}} & GDP-TabICL & 1 & 0.77\textsubscript{.02} & 0.76\textsubscript{.00} & 0.77\textsubscript{.01} & 0.77\textsubscript{.02} & 0.77\textsubscript{.01} & 0.77\textsubscript{.01} & 0.76\textsubscript{.01}\\
  & GDP-TabICL & 2 & \textbf{0.83\textsubscript{.01}} & 0.81\textsubscript{.03} & \textbf{0.82\textsubscript{.01}} & 0.80\textsubscript{.02} & 0.81\textsubscript{.02} & 0.78\textsubscript{.01} & 0.80\textsubscript{.03}\\
  & GDP-TabICL & 4 & 0.81\textsubscript{.01} & 0.81\textsubscript{.02} & 0.80\textsubscript{.01} & 0.81\textsubscript{.01} & 0.80\textsubscript{.01} & 0.80\textsubscript{.01} & 0.80\textsubscript{.01}\\
  & GDP-TabICL & 8 & 0.80\textsubscript{.00} & \textbf{0.82\textsubscript{.02}} & \textbf{0.82\textsubscript{.02}} & \textbf{0.82\textsubscript{.01}} & \textbf{0.82\textsubscript{.01}} & \textbf{0.83\textsubscript{.01}} & \textbf{0.83\textsubscript{.01}}\\
 & Diffprivlib LR & - & 0.43\textsubscript{.23}& 0.52\textsubscript{.22}& 0.36\textsubscript{.22}& 0.50\textsubscript{.01}& 0.51\textsubscript{.24}& 0.63\textsubscript{.23}& \textbf{0.83\textsubscript{.01}}\\ 
 & Diffprivlib GNB & - & 0.69\textsubscript{.08}& 0.59\textsubscript{.19}& 0.67\textsubscript{.13}& 0.73\textsubscript{.02}& 0.74\textsubscript{.03}& 0.76\textsubscript{.03}& 0.78\textsubscript{.03}\\ \midrule
 
\multirow{6}{*}{\underline{Bank}}  & GDP-TabICL & 1 & \textbf{0.89\textsubscript{.01}} & 0.88\textsubscript{.01} & \textbf{0.88\textsubscript{.01}} & 0.88\textsubscript{.01} & \textbf{0.88\textsubscript{.01}} & 0.88\textsubscript{.01} & 0.88\textsubscript{.01}\\
  & GDP-TabICL & 2 & 0.88\textsubscript{.01} & \textbf{0.89\textsubscript{.01}} & \textbf{0.88\textsubscript{.01}} & \textbf{0.89\textsubscript{.01}} & \textbf{0.88\textsubscript{.00}} & 0.88\textsubscript{.01} & \textbf{0.89\textsubscript{.01}}\\
  & GDP-TabICL & 4 & 0.88\textsubscript{.01} & 0.88\textsubscript{.02} & 0.87\textsubscript{.01} & 0.88\textsubscript{.01} & \textbf{0.88\textsubscript{.02}} & 0.88\textsubscript{.01} & 0.88\textsubscript{.01}\\
  & GDP-TabICL & 8 & 0.88\textsubscript{.01} & \textbf{0.89\textsubscript{.01}} & \textbf{0.88\textsubscript{.01}} & 0.88\textsubscript{.01} & \textbf{0.88\textsubscript{.02}} &\textbf{0.89\textsubscript{.01}} & 0.88\textsubscript{.00}\\
& Diffprivlib LR & - & 0.47\textsubscript{.26}& 0.41\textsubscript{.25}& 0.49\textsubscript{.03}& 0.52\textsubscript{.26}& 0.39\textsubscript{.21}& 0.50\textsubscript{.38}& \textbf{0.89\textsubscript{.01}}\\ 
 & Diffprivlib GNB & - & 0.84\textsubscript{.07}& 0.78\textsubscript{.12}& 0.83\textsubscript{.04}& 0.85\textsubscript{.02}& 0.84\textsubscript{.03}& 0.83\textsubscript{.02}& 0.82\textsubscript{.03} \\ \midrule
 
 \multirow{6}{*}{\underline{Blood}} & GDP-TabICL & 1 & 0.58\textsubscript{.18} & 0.65\textsubscript{.04} & 0.65\textsubscript{.03} & 0.66\textsubscript{.01} & 0.63\textsubscript{.01} & 0.64\textsubscript{.04} & 0.64\textsubscript{.05}\\
  & GDP-TabICL & 2 & \textbf{0.79\textsubscript{.02}} & \textbf{0.81\textsubscript{.02}} & \textbf{0.78\textsubscript{.02}} & \textbf{0.80\textsubscript{.00}} & \textbf{0.78\textsubscript{.01}} & 0.79\textsubscript{.02} & 0.78\textsubscript{.02}\\
  & GDP-TabICL & 4 & 0.64\textsubscript{.22} & 0.68\textsubscript{.14} & 0.73\textsubscript{.09} & 0.76\textsubscript{.02} & 0.76\textsubscript{.02} & 0.75\textsubscript{.03} & 0.75\textsubscript{.04}\\
  & GDP-TabICL & 8 & 0.21\textsubscript{.03} & 0.22\textsubscript{.04} & 0.37\textsubscript{.12} & 0.40\textsubscript{.14} & 0.23\textsubscript{.02} & 0.23\textsubscript{.10} & 0.32\textsubscript{.10}\\
 & Diffprivlib LR & - & 0.43\textsubscript{.28}& 0.45\textsubscript{.27}& 0.43\textsubscript{.27}& 0.57\textsubscript{.28}& 0.52\textsubscript{.28}& 0.51\textsubscript{.30}& \textbf{0.83\textsubscript{.03}}\\ 
 & Diffprivlib GNB & - & 0.77\textsubscript{.08}& 0.75\textsubscript{.11}& \textbf{0.78\textsubscript{.03}}& 0.78\textsubscript{.02}& 0.76\textsubscript{.08}& \textbf{0.80\textsubscript{.02}}& 0.82\textsubscript{.03}\\ \midrule
  
 \multirow{6}{*}{\underline{Car}}  & GDP-TabICL &  1 & 0.75\textsubscript{.06} & 0.70\textsubscript{.09} & 0.67\textsubscript{.13} & 0.74\textsubscript{.05} & 0.73\textsubscript{.07} & 0.73\textsubscript{.08} & 0.69\textsubscript{.09}\\
  & GDP-TabICL & 2 & \textbf{0.76\textsubscript{.03}} & 0.72\textsubscript{.02} & 0.74\textsubscript{.03} & 0.74\textsubscript{.02} & 0.75\textsubscript{.02} & 0.75\textsubscript{.03} & 0.75\textsubscript{.02}\\
  & GDP-TabICL & 4 & \textbf{0.76\textsubscript{.04}} & \textbf{ 0.75\textsubscript{.02}} & 0.77\textsubscript{.01} & 0.78\textsubscript{.02} & 0.77\textsubscript{.01} & 0.75\textsubscript{.02} & 0.77\textsubscript{.02}\\
  & GDP-TabICL & 8 & 0.75\textsubscript{.03} & 0.69\textsubscript{.06} & 0.75\textsubscript{.04} & 0.73\textsubscript{.04} & 0.75\textsubscript{.03} & 0.74\textsubscript{.05} & 0.75\textsubscript{.02}\\
& Diffprivlib LR & - & 0.46\textsubscript{.15}& 0.52\textsubscript{.13}& 0.56\textsubscript{.10}& 0.52\textsubscript{.10}& 0.58\textsubscript{.18}&0.75\textsubscript{.06} & \textbf{0.94\textsubscript{.01}}\\ 
 & Diffprivlib GNB & - & 0.66\textsubscript{.09}& 0.72\textsubscript{.13}& \textbf{0.84\textsubscript{.08}}& \textbf{0.82\textsubscript{.09}}& \textbf{0.89\textsubscript{.02}}& \textbf{0.88\textsubscript{.02}}& 0.85\textsubscript{.01}\\ \midrule
 
 \multirow{6}{*}{Calhousing}  & GDP-TabICL & 1 & 0.56\textsubscript{.04} & 0.55\textsubscript{.05} & 0.56\textsubscript{.03} & 0.54\textsubscript{.06} & 0.52\textsubscript{.07} & 0.56\textsubscript{.06} & 0.54\textsubscript{.06}\\
  & GDP-TabICL & 2 & 0.53\textsubscript{.06} & 0.49\textsubscript{.02} & 0.54\textsubscript{.04} & 0.49\textsubscript{.03} & 0.48\textsubscript{.04} & 0.49\textsubscript{.03} & 0.48\textsubscript{.01}\\
  & GDP-TabICL & 4 & 0.52\textsubscript{.05} & 0.56\textsubscript{.06} & 0.57\textsubscript{.05} & 0.53\textsubscript{.03} & 0.55\textsubscript{.04} & 0.53\textsubscript{.08} & 0.63\textsubscript{.03}\\
  & GDP-TabICL & 8 & 0.53\textsubscript{.05} & 0.56\textsubscript{.06} & 0.53\textsubscript{.04} & 0.57\textsubscript{.07} & 0.57\textsubscript{.09} & 0.61\textsubscript{.08} & 0.61\textsubscript{.03}\\
& Diffprivlib LR & - & 0.50\textsubscript{.06}& 0.53\textsubscript{.06}& 0.47\textsubscript{.04}& 0.50\textsubscript{.05}& 0.51\textsubscript{.05}& 0.52\textsubscript{.02}& \textbf{0.77\textsubscript{.03}}\\ 
 & Diffprivlib GNB & - & \textbf{0.57\textsubscript{.11}}& 0.58\textsubscript{.08}& \textbf{0.69\textsubscript{.07}}& \textbf{0.69\textsubscript{.07}}& \textbf{0.69\textsubscript{.06}}& \textbf{0.70\textsubscript{.06}}& 0.72\textsubscript{.02}\\ \midrule
 
 \multirow{6}{*}{Diabetes}  & GDP-TabICL & 1 & 0.64\textsubscript{.01} & 0.66\textsubscript{.02} & 0.66\textsubscript{.03} & 0.65\textsubscript{.01} & 0.65\textsubscript{.02} & 0.65\textsubscript{.01} & 0.65\textsubscript{.02}\\
  & GDP-TabICL & 2 & \textbf{0.67\textsubscript{.02}} & 0.65\textsubscript{.02} &  0.65\textsubscript{.02} &  0.69\textsubscript{.01} &  0.68\textsubscript{.02} &  0.67\textsubscript{.02} &  0.66\textsubscript{.02}\\
  & GDP-TabICL & 4 &  \textbf{0.67\textsubscript{.02}} &  0.68\textsubscript{.03} & 0.68\textsubscript{.03} & \textbf{0.71\textsubscript{.02}} & 0.70\textsubscript{.01} & 0.72\textsubscript{.03} & 0.70\textsubscript{.03}\\
  & GDP-TabICL & 8 & 0.45\textsubscript{.13} & 0.48\textsubscript{.13} & 0.56\textsubscript{.12} & 0.65\textsubscript{.09} & 0.70\textsubscript{.04} & 0.67\textsubscript{.04} & 0.71\textsubscript{.03}\\
& Diffprivlib LR & - & 0.42\textsubscript{.13}& 0.51\textsubscript{.16}& 0.54\textsubscript{.14}& 0.42\textsubscript{.11}& 0.47\textsubscript{.15}& 0.45\textsubscript{.17}& 0.77\textsubscript{.01}\\ 
 & Diffprivlib GNB & - & 0.61\textsubscript{.08}& \textbf{0.69\textsubscript{.04}}& \textbf{0.71\textsubscript{.05}}& 0.68\textsubscript{.08}& \textbf{0.78\textsubscript{.03}}& \textbf{0.77\textsubscript{.03}}& \textbf{0.79\textsubscript{.02}}\\ \midrule
 
 \multirow{6}{*}{Heart}  & GDP-TabICL & 1 & 0.56\textsubscript{.06} & 0.53\textsubscript{.03} & 0.53\textsubscript{.01} & 0.55\textsubscript{.03} & 0.54\textsubscript{.01} & 0.57\textsubscript{.02} & 0.54\textsubscript{.02}\\
  & GDP-TabICL & 2 & 0.51\textsubscript{.03} & 0.49\textsubscript{.03} & 0.50\textsubscript{.02} & 0.50\textsubscript{.02} & 0.50\textsubscript{.03} & 0.49\textsubscript{.01} & 0.49\textsubscript{.03} \\
  & GDP-TabICL & 4 & 0.59\textsubscript{.06} & 0.65\textsubscript{.03} & 0.70\textsubscript{.03} & 0.70\textsubscript{.04} & 0.66\textsubscript{.06} & 0.67\textsubscript{.03} & 0.70\textsubscript{.01}\\
  & GDP-TabICL & 8 & 0.56\textsubscript{.08} & 0.62\textsubscript{.07} & 0.56\textsubscript{.05} & 0.55\textsubscript{.04} & 0.54\textsubscript{.05} & 0.56\textsubscript{.02} & 0.53\textsubscript{.05}\\
& Diffprivlib LR & - & 0.47\textsubscript{.06}&0.49\textsubscript{.07}& 0.50\textsubscript{.08}& 0.48\textsubscript{.07}& 0.54\textsubscript{.08}& 0.54\textsubscript{.05}& \textbf{0.86\textsubscript{.03}}\\ 
 & Diffprivlib GNB & - & \textbf{0.74\textsubscript{.08}}& 0.74\textsubscript{.11}& \textbf{0.76\textsubscript{.10}}& \textbf{0.78\textsubscript{.09}}& \textbf{0.84\textsubscript{.02}}& \textbf{0.85\textsubscript{.02}}& 0.84\textsubscript{.02}\\ \midrule
 
 \multirow{6}{*}{Jungle}  & GDP-TabICL & 1 & 0.47\textsubscript{.02} & 0.48\textsubscript{.01} & 0.49\textsubscript{.01} & 0.48\textsubscript{.02} & 0.48\textsubscript{.02} & 0.48\textsubscript{.00} & 0.48\textsubscript{.02} \\
  & GDP-TabICL & 2 & 0.49\textsubscript{.02} & 0.49\textsubscript{.01} & 0.48\textsubscript{.01} & 0.48\textsubscript{.01} & 0.48\textsubscript{.01} & 0.49\textsubscript{.02} & 0.48\textsubscript{.01}\\
  & GDP-TabICL & 4 & 0.50\textsubscript{.01} & 0.48\textsubscript{.01} & 0.48\textsubscript{.03} & 0.48\textsubscript{.01} & 0.49\textsubscript{.02} & 0.48\textsubscript{.01} & 0.49\textsubscript{.02}\\
  & GDP-TabICL & 8 & 0.49\textsubscript{.01} & 0.48\textsubscript{.01} & 0.49\textsubscript{.01} & 0.48\textsubscript{.01} & 0.47\textsubscript{.04} & 0.48\textsubscript{.01} & 0.49\textsubscript{.03}\\
& Diffprivlib LR & - & 0.50\textsubscript{.05}& 0.51\textsubscript{.04}& 0.45\textsubscript{.05}& 0.52\textsubscript{.06}& 0.52\textsubscript{.02}& 0.73\textsubscript{.02}& 0.73\textsubscript{.01}\\ 
 & Diffprivlib GNB & - & \textbf{0.70\textsubscript{.07}}& \textbf{0.71\textsubscript{.08}}& \textbf{0.75\textsubscript{.01}}& \textbf{0.75\textsubscript{.02}}& \textbf{0.75\textsubscript{.01}}& \textbf{0.75\textsubscript{.01}}& \textbf{0.75\textsubscript{.01}}\\ 
 \bottomrule
\end{tabular}%
}
\end{table}

\end{appendix}
\end{document}